\documentclass[fleqn,usenatbib,useAMS]{mnras}

\usepackage{graphicx}	
\usepackage{wrapfig}
\usepackage{lscape}
\usepackage{rotating}
\usepackage{amsmath}	
\usepackage{amssymb}	
\usepackage{multicol}        
\usepackage{bm}		
\usepackage{pdflscape}	
\usepackage{natbib}
\usepackage{hyperref}
\usepackage{multirow}
\usepackage{arydshln}
\usepackage{float}
\usepackage{subfigure}
\usepackage{lineno}

\usepackage[T1]{fontenc}
\usepackage{ae,aecompl}

\title[]{Evolution of the dusty nova QY Mus from eruption to quiescence}
 \author[M. S. Bisht et al.]{
 Mohit Singh Bisht$^{1}$,
 A. Raj$^{1}$\thanks{E-mail: ashishpink@gmail.com },
 F.M. Walter$^{2}$,
 D. Bisht$^{1}$,
 K. Belwal$^{1}$,
 \newauthor
 S. Biswas$^{1}$
 \\
 $^{1}$Indian Centre for Space Physics, 466 Barakhola, Netai Nagar, Kolkata 700099, West Bengal, India\\
 $^{2}$Department of Physics and Astronomy, Stony Brook University, Stony Brook, NY 11794-3800, USA\\
 }

\date{Last updated XXX ; in original form YYY}
\pubyear{2026}

\begin{document}
\label{firstpage}
\pagerange{\pageref{firstpage}--\pageref{lastpage}}
\maketitle

\begin{abstract}
We present a comprehensive study of the spectrophotometric evolution of the classical nova QY Mus from eruption to quiescence. The light curve shows a notable dust dip, classifying it as a D (137)-type nova, with dust formation beginning at $\sim$123 days post-outburst and reaching a maximum optical depth of $\tau \sim 3.2$. We classify QY Mus as a slow nova with $t_2 = 87 \pm 6$ days, and derive an absolute magnitude of $M_V = -6.55 \pm 0.54$ using the MMRD relation.
The spectroscopic evolution, traced from 94 to 1348 days, shows prominent P-Cygni profiles in Balmer and Fe\,\textsc{ii} lines during the early decline, consistent with an Fe\,\textsc{ii}-type nova. The transition to the nebular phase occurs around $\sim$233 days, marked by the emergence of [O\,\textsc{iii}] emission. Photoionization modeling using \textsc{Cloudy} of 41 emission lines on day 590 yields a central source temperature of $(7.08 \pm 0.20)\times10^{5}$ K, with enhanced nitrogen and oxygen abundances and moderate neon enrichment, suggesting that QY~Mus is not a neon nova.
Mid-infrared WISE observations at $\sim$502 days indicate the presence of cool dust at $\sim$400~K. Using a Gaia-based color--magnitude diagram constructed in this work for 34 quiescent novae, we find that QY~Mus occupies a region consistent with systems hosting main-sequence or subgiant secondaries; its orbital period further supports a subgiant companion. These results establish QY~Mus as a slow, dust-forming nova with well-characterized evolution and a subgiant secondary.
\end{abstract}

\begin{keywords}
cataclysmic variables --- classical novae --- QY Mus---- techniques : spectroscopic -- line : identification -- dusty nova
\end{keywords}




\section{Introduction} \label{sec:intro}
 
Novae are thermonuclear eruptions occurring on the surface of a white dwarf (WD) in a close binary system.  In these systems, the WD accretes hydrogen-rich material from a nearby donor star, i.e., the secondary star. The material forms an accretion disk around the WD before being transferred to the surface of the white dwarf. The accreted hydrogen-rich material on the surface of the white dwarf undergoes a thermonuclear runaway, leading to a sudden increase in luminosity by several orders of magnitude and the ejection of material at velocities ranging from a few hundred to several thousand km\,s$^{-1}$\cite[][and references therein]{BodeEvansBook2008, Jose2020, Starrfield2020, chomiuk2021new}. The secondary stars in nova systems are classified as main-sequence (MS-novae), subgiant (SG-novae), and red giant (RG-novae) \citep{2012ApJ...746...61D}. MS-novae typically have short orbital periods ranging from $1.4$ to 10 hours \citep{1995cvs..book.....W, 1997A&A...322..807D, chomiuk2021new},
while SG-novae host moderately evolved secondaries and have orbital periods of $0.6$--10~days. In contrast, RG-novae host giant companions and are characterized by long orbital periods, typically  10–1000 days \citep{schaefer2022comprehensive}.

Nova QY Mus / Nova Mus 2008 was discovered during eruption on 2008 September 28.99 UT by \cite{2008IAUC.8990....2L} with an apparent magnitude of $\sim$8.6 on a pair of Technical Pan photographs taken with an 85-mm camera lens. Additional magnitudes reported by Liller show that the nova was below the limiting magnitude (unfiltered 11.5) on September 15.02 UT.
Further observations by V. Tabur, reported in \cite{2008IAUC.8990....2L}, present unfiltered CCD measurements, yielding a position of R.A. = 13h16m36s.22, Decl. = $-67^\circ36'50''.7$ (uncertainty $\pm$ 4$''$). The reported photometry indicates that the nova was below the  limiting magnitude (unfiltered 11.6) on September 19.42 UT.
The magnitudes reported by Liller and V. Tabur clearly indicate that the nova was below the unfiltered limiting magnitude on September 15.02 (11.5) and 19.42 UT (11.6) and had brightened   
to magnitude 9.9 by September 21.38 UT. 
The magnitudes in $B$ and $V$ were 10.26 and 9.31 on October 2.44 UT, with position end figures 36s.44, 47$''$.8, as reported in the \cite{2008IAUC.8990....2L}. A. Skiff (Lowell Observatory) reported that the archival position matches a USNO-B1.0 catalogue star with blue magnitude 19.9 and red magnitude 17.5, having position end figures 36s.47, 47$''$.9. A star is also present within 0$''$.5 of the nova position on a plate taken via the U.K. Schmidt Telescope H$\alpha$ survey.

An underexposed spectrum obtained by \cite{2008IAUC.8990....2L} on October 4.01 UT with a Schmidt camera  showed a single broad emission feature at the position of H$\alpha$, with a width of approximately 230 nm. 
Additional spectroscopic observations conducted on 2008 December 24.3 UT by F.~M.~Walter, using the SMARTS 1.5-m telescope, classified the object as a typical Fe~II-type nova \citep{2009IAUC.9012....2L}. The resulting spectrum was characterized by prominent Balmer emission lines,
with H$\alpha$ exhibiting a FWHM of $\sim$3.2 nm and a clearly defined blueshifted absorption component at $-1650$ km~s$^{-1}$, along with the presence of [O~I] 6300, 6364~\AA. Follow-up spectra on December 28.3 UT showed strong Balmer emission lines with deep wind absorption features, while Fe~II multiplets remained prominent, several displaying blueshifted P-Cygni absorption profiles.

In this work, we present a comprehensive analysis of the poorly studied dusty nova QY Mus. The structure of the paper is as follows. The observations are described in Section~\ref{observations}.
Section~\ref{analysis} presents the evolution of the optical and NIR light curves from maximum brightness to quiescence, including the dust-dip phase. The detailed optical spectroscopic evolution during the early decline and nebular stages is also discussed. Photoionization modeling is performed using the \texttt{CLOUDY} code to derive the physical and chemical parameters of the system. An analysis of the dust properties is carried out using data from the Wide-field Infrared Survey Explorer (WISE).
The possible progenitor of QY~Mus is discussed in Section~\ref{Discussion}. Using Gaia DR3 photometric data, we construct a color–magnitude diagram and compare the position of QY~Mus with that of known quiescent novae. We find that QY~Mus occupies a region consistent with systems hosting main-sequence and subgiant secondaries. A summary is presented in Section \ref{summary}.

\section{Observations}\label{observations}
\subsection{Optical and NIR photometry}

All photometric data used in this study were obtained from several public databases, including the American Association of Variable Star Observers (AAVSO\footnote{\href{https://www.aavso.org/}{https://www.aavso.org/}}) international database. We obtained $B$ and $V$ band data from AAVSO.
We obtained $B$, $V$, $R_{\mathrm{C}}$, and $I_{\mathrm{C}}$ photometric data for QY~Mus from the Variable Star Observers League in Japan database (VSOLJ\footnote{\href{http://vsolj.cetus-net.org/}{http://vsolj.cetus-net.org/}}).
 We also retrieved $V$-band data from the All Sky Automated Survey (ASAS\footnote{\href{https://www.astrouw.edu.pl/asas/?page=aasc\&catsrc=asas3/}{https://www.astrouw.edu.pl/asas/?page=aasc\&catsrc=asas3/}}; \cite{1997AcA....47..467P}).

 Additional optical and near-infrared (NIR) data were obtained from the Small and Moderate Aperture Research Telescope System (SMARTS\footnote{\href{http://www.astro.sunysb.edu/fwalter/SMARTS/NovaAtlas/}{http://www.astro.sunysb.edu/fwalter/SMARTS/NovaAtlas/}}). These data were taken from the Stony Brook/SMARTS Atlas of (mostly) Southern Novae, which has provided spectra and photometry for novae since 2003. This atlas enables systematic studies of nova behaviour as well as detailed investigations of individual objects. 

\subsection{WISE data}

Mid-infrared (mid-IR) band data were obtained from the WISE catalogue\footnote{\href{https://irsa.ipac.caltech.edu}{https://irsa.ipac.caltech.edu}}
and used in this study to derive the parameters associated with dust formation. 
WISE is a space-based infrared survey mission that scanned the sky in the four bands: 
W1 (3.35\,$\mu$m), W2 (4.60\,$\mu$m), W3 (11.56\,$\mu$m), and W4 (22.09\,$\mu$m) \citep{wright2010wide}.

\subsection{Spectroscopy}
In this study, low-dispersion optical spectra were obtained from the publicly available SMARTS Nova Atlas. The observations were carried out with an irregular cadence, covering the early decline and nebular phases of the nova evolution. These spectra were collected under various sky conditions. The SMARTS RC grating spectrograph was used to obtain the spectra. 
Details on the instrument setup, data collection strategies, and reduction steps are provided in \cite{wal12}, which also includes a brief discussion of QY~Mus, identifying it as a typical Fe~II nova that underwent dust formation. In total, 35 spectra were obtained, spanning approximately 3.7 years, from 2008 December 24 (day 94) to 2012 June 1 (day 1348). 
The details of the spectroscopic observations are summarized in Table~\ref{log}.

\begin{table}
\centering
\caption{Observational log for spectroscopic data obtained for QY Mus.}
\label{log}
\resizebox{\hsize}{!}{%
\begin{tabular}{lccccc}
\hline
\hline
& \textbf{Time since}    & \textbf{Wavelength} & \textbf{Resolution} \\
\textbf{Date (UT)} & \textbf{discovery}   & \textbf{range} & \textbf{(\AA)} \\
& \textbf{(days)} & \textbf{(\AA)} & &   \\
\hline
			2008 December 24.28 & 94   & 5629--6949 & 3.1 \\[0.25ex]
            2008 December 28.36 & 98   & 3651--5420 & 4.1 \\[0.25ex]
			2009 March 28.04 & 187.66   & 3646--5415 & 4.1  \\[0.25ex]
            2009 March 30.27 & 189.89  & 3644--5414 & 4.1  \\[0.25ex]
			2009 April 04.04 & 194.66  & 3647--5416 & 4.1  \\[0.25ex]
            2009 April 05.34 & 196  & 5622--6941 & 3.1  \\[0.25ex]
            2009 April 15.08 & 205.70  & 3648--5417 & 4.1  \\[0.25ex]
            2009 April 17.12 & 207.74  & 5626--6943 & 3.1  \\[0.25ex]
			2009 May 01.05 & 221.66  & 3649--5415 & 4.1 \\[0.25ex]
            2009 May 13.03 & 233.64  & 3648--5417 & 4.1 \\[0.25ex]
            2009 May 27.00 & 247.46  & 3650--5417 & 4.1 \\[0.25ex]
			2009 June 01.08 & 252.70 & 3650--5415 & 4.1  \\[0.25ex]
			2009 July 02.11 & 283.74  & 3869--4542 & 1.6 \\[0.25ex]
			2009 July 20.05 & 301.68   & 5627--6946 & 3.1  \\[0.25ex]
            2009 September 03.97 & 347.60    & 4062--4735 & 1.6   \\[0.25ex]
            2009 September 10.98 & 354.60   & 5627--6945 & 3.1   \\[0.25ex]
            2009 September 13.98 & 357.60    & 3646--5414 & 4.1   \\[0.25ex]
            2009 December 20.32 & 454.99   & 5628--6946  & 3.1 \\[0.25ex]
            2009 December 21.30 & 455.91    & 3651--5420  & 4.1 \\[0.25ex]
            2010 January  15.22 & 480.84    & 3646--5416 & 4.1  \\[0.25ex]
            2010 March 09.27 & 533.89   & 3652--5420 & 4.1  \\[0.25ex]
            2010 March 24.34 & 548.96   & 5626--6945 & 3.1  \\[0.25ex]
            2010 May 05.26 & 590.88  & 3240--9563 & 17.2  \\[0.25ex]
            2010 June 02.01 & 618.63 & 3264--9539 & 17.2  \\[0.25ex]
            2010 December 15.32 & 814.94    & 5628--6944  & 3.1 \\[0.25ex]
            2011 February 16.24 & 877.86    & 3238--9530 & 17.2  \\[0.25ex]
            2011 March 19.20 & 908.82  & 5623--6942 & 3.1  \\[0.25ex]
            2011 August 22.00 & 1064.62    & 3271--9604 & 17.2 \\[0.25ex]
            2011 December 23.32 & 1187.94   & 5622--6941  & 3.1 \\[0.25ex]
            2011 December 30.33 & 1194.95   & 3256--9605  & 17.2\\[0.25ex]
            2012 January  11.27 & 1206.89   & 3645--5415 & 4.1  \\[0.25ex]
            2012 February 03.22 & 1229.84   & 3253--9529 & 17.2  \\[0.25ex]
            2012 February 05.27 & 1231.89   & 5627--6945 & 3.1  \\[0.25ex]
            2012 February 18.28 & 1244.90   & 3646--5416 & 4.1  \\[0.25ex]
            2012 June 01.12 & 1348.74   & 3652--5420 & 4.1 \\[0.25ex]
\hline
\end{tabular}}
\end{table}

\section{Analysis}\label{analysis}

\subsection{Optical light curve and photometric parameters}\label{optical light curve}
The optical light curves based on the data from the AAVSO, SMARTS, ASAS, and VSOLJ observation database are presented in Fig. \ref{lc_optical}. The nova was below the limiting magnitude (11.6) on 2008 September 19.42 UT and had brightened to magnitude 9.9 (unfiltered) by 2008 September 21.38 UT \citep{2008IAUC.8990....2L}. 
In this paper, we adopt 2008 September 21.38 UT (JD 2454730.878) as day 0, corresponding to the outburst date. The $B$, $V$, $R$, and $I$ photometric coverage starts from day 11.
The brightness reached the peak magnitude $B_{\mathrm{max}}$ = 9.38, $V_{\mathrm{max}}$ =8.61, $R_{\mathrm{max}}$= 8.08, $I_{\mathrm{max}}$ =7.44 on day 18.06. The post-maximum evolution of QY~Mus indicates a slow decline. The classification of nova speed relies on the parameter $t_2$, defined as the days required to drop by 2 magnitudes from peak brightness. We performed a least-squares fit to the $V$-band data and found $ t_2 = 87 \pm 6$ days, placing the system in the slow-nova category \citep{mclaughlin1945relation,1957gano.book.....G}. Due to the dust-dip observed in the light curve, the $t_3$ value was estimated from the empirical relation $t_3 = 2.75\, t_2^{0.88}$ \citep{1995cvs..book.....W}, giving $t_3 = 137 \pm 8$ days.
The $t_3$ value obtained from the relation is consistent with the value derived by extrapolating the $t_2$ fit of the $V$-band light curve to a decline of 3 mag from maximum.

We apply the recent maximum magnitude versus rate of decline (MMRD) relation by \cite{2019A&A...622A.186S}, which is based on parallaxes from \textit{Gaia} DR2. Utilizing this relation, we estimate a maximum absolute magnitude of $M_V = -6.55 \pm 0.54$ using the $t_3$ parameter in the $V$ band.
This corresponds to a distance modulus of $V_{\rm max} - M_V = 15.16$~mag (see Section~\ref{reddening and distance} for the distance estimation).
From relation given by \cite{livio1992classical} which uses maximum  absolute magnitude M$_V$ , $t_3$ to derive the  mass of the white dwarf (M$_{WD}$), for nova QY Mus we determined the mass of underlying white dwarf is 0.70 $\pm$ 0.03 M$_\odot$. 
This suggests that nova QY~Mus hosts a low-mass carbon-oxygen (CO) type WD ($M_{\rm WD} \leq 1.2\,M_\odot$). Based on $M_V = -6.55$, the outburst luminosity is calculated as $(4.24 \pm 0.30)\times10^{4}\,L_\odot$.

\subsection{Dust dip}\label{dust dip}
The BVRI light curves display a gradual decline in brightness following the maximum, continuing up to day 117. After day 121, the brightness decay rate increases significantly.
The decay rate in $V$ band was about $0.022 \pm 0.003$ mag~day$^{-1}$ until day 121, and changed to $0.29 \pm 0.06$ mag~day$^{-1}$ between days 121 and 124. This increased decay rate is indicative of a dust dip, indicating obscuration of the optical flux by newly formed dust in the nova ejecta. This dip in optical light curve characteristic of dust is supported by NIR observations from SMARTS data, which display a rise across the $J$, $H$, and $K$ bands (see Section~\ref{nir_lc} for more description). The dust formation phase is also clearly reflected in the color--magnitude diagram (CMD), where the nova exhibits a pronounced shift toward redder $(B - V)_0$ colors along with a rapid decline in luminosity during this phase (see Fig.~\ref{cmd_track}). The CMD track shows a deviation from the earlier evolutionary path, moving to the right and downward. 

The brightnesses in the $B$, $V$, $R$, and $I$ bands were 11.77, 11.05, 9.72, and 10.28 mag, respectively, on day 117, and decreased to 13.98, 13.13, 11.85, and 11.19 mag within day 124. The $V$ band mag from ASAS was 11.17 mag on day 121. 
No data were obtained between days 121 and 124; therefore, the onset of the dust dip is considered to occur at approximately $123 \pm 2$ days after the outburst. The $V$, $R$, and $I$ bands reached their minimum during the dust dip on day 146.
After the minima, all bands began to recover, and by day $\sim$200 they had fully recovered. 
Following day 200, the photometric data indicate a progressive fading, entering the final decline phase of the light curve.
The system entered into quiescence after $\sim$2500 days, as indicated by its position in the CMD (Fig.~\ref{cmd_track}), where it occupies a region similar to other quiescence novae \citep[see their Fig.~1]{2012ApJ...746...61D}.

We fit a polynomial to the $V$-band light curve starting from $t_{\rm peak}$. 
The visual optical depth ($\tau$) of the dust shell is calculated as the difference in magnitude between the observed values and those predicted by the fitted function. Figure~\ref{depth} presents the evolution of $\tau$ following maximum brightness. The maximum value, $\tau_{\rm max} \sim 3.2$, reached on day 146.
From Figure~\ref{depth}, it is clear that the onset of dust formation occurred around $123 \pm 2$ days and that the nova recovered after $\sim$200 days since discovery.

\begin{figure*}
	\includegraphics[scale=0.55]{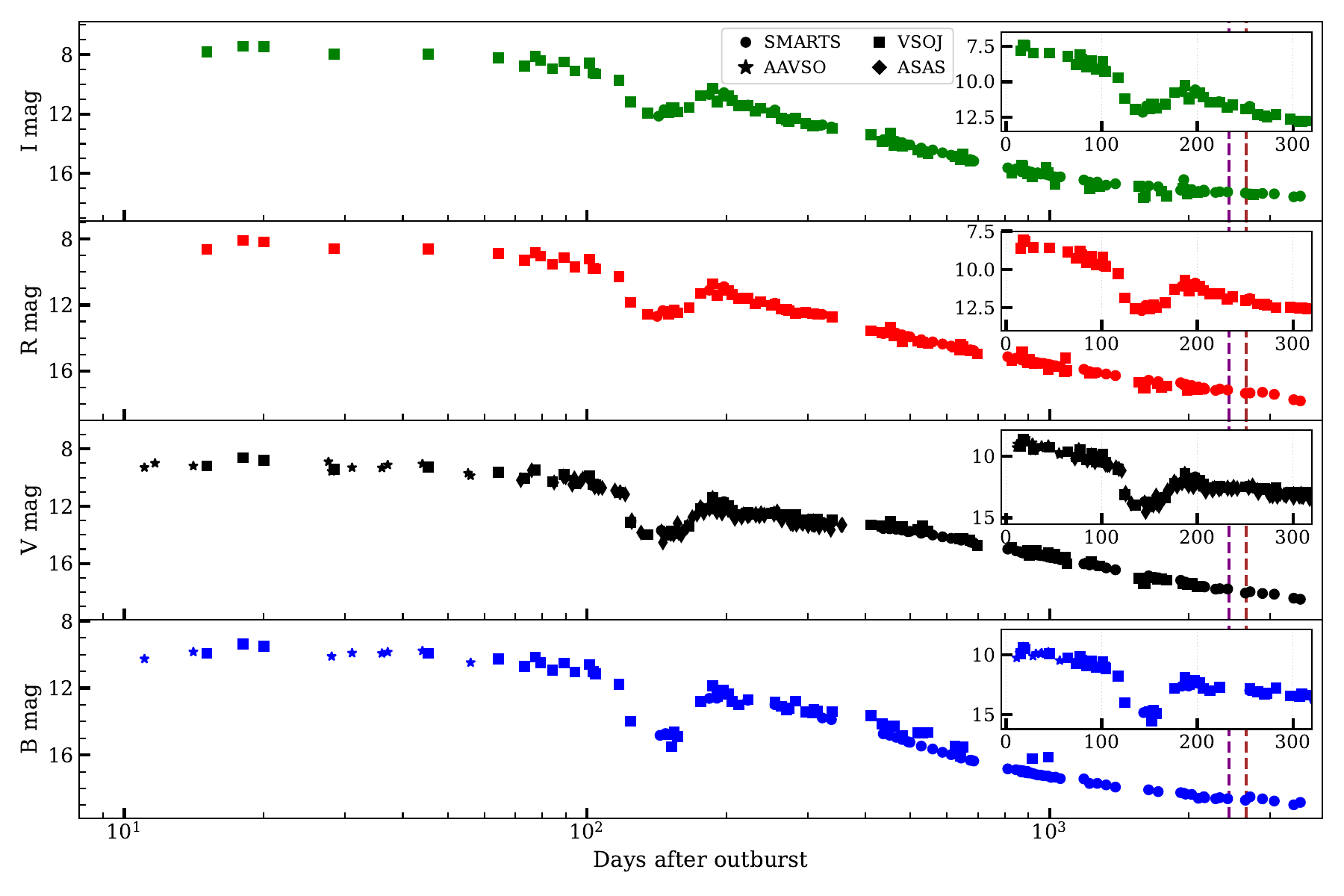}
\caption{The light curve evolution of QY~Mus over days 11 to 3482 is constructed from observations obtained with the SMARTS, AAVSO, VSOLJ, and ASAS databases. The inset panel presents the first 310 days in detail, highlighting the optical dip due to obscuration of the optical flux by dust.
The purple and brown dashed lines denote the Gaia observation epochs corresponding to DR2 (2015.5; \citealt{2018A&A...616A...1G}) and DR3 (2016.0; \citealt{2021A&A...649A...1G}), respectively.
}
	\label{lc_optical}
\end{figure*}

\begin{figure}
    \centering
    \includegraphics[width=1.05\linewidth]{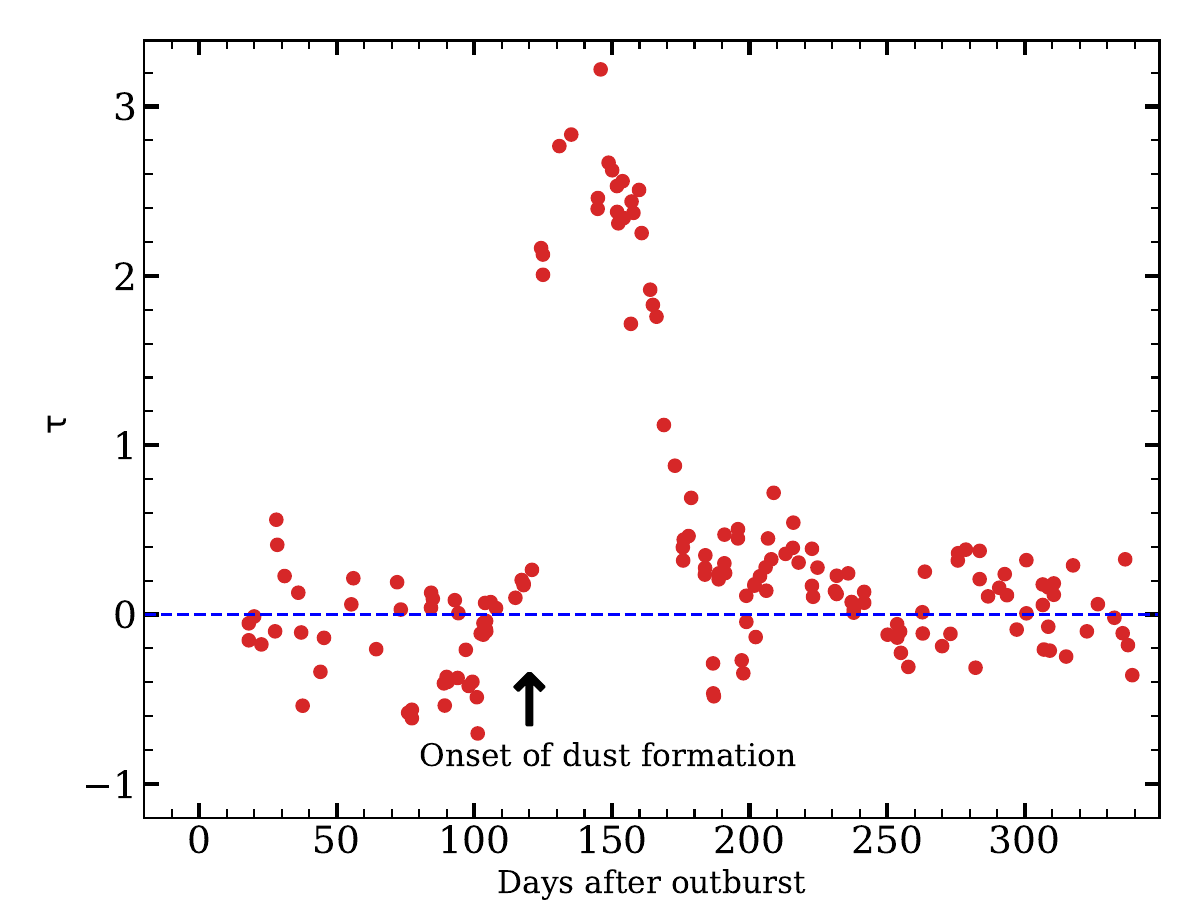}
    \caption{Visual light curve derived from the difference between the observed magnitude and the fitted magnitude prior to the onset of dust formation, showing the temporal evolution of the visual optical depth. }
    \label{depth}
\end{figure}

Following the classification of nova light-curve shapes and decline time ($t_3$) given by \cite{strope2010catalog}, the optical light curve of QY~Mus, as shown in Fig.~\ref{lc_optical}, exhibits a characteristic dust dip, classifying it as a D(137)-class nova. This dust dip is well covered by high-cadence photometric observations from SMARTS, VSOLJ, and ASAS.

\subsection{Evolution of optical colors}\label{optical color}

\begin{figure}
	\includegraphics[scale=0.42]{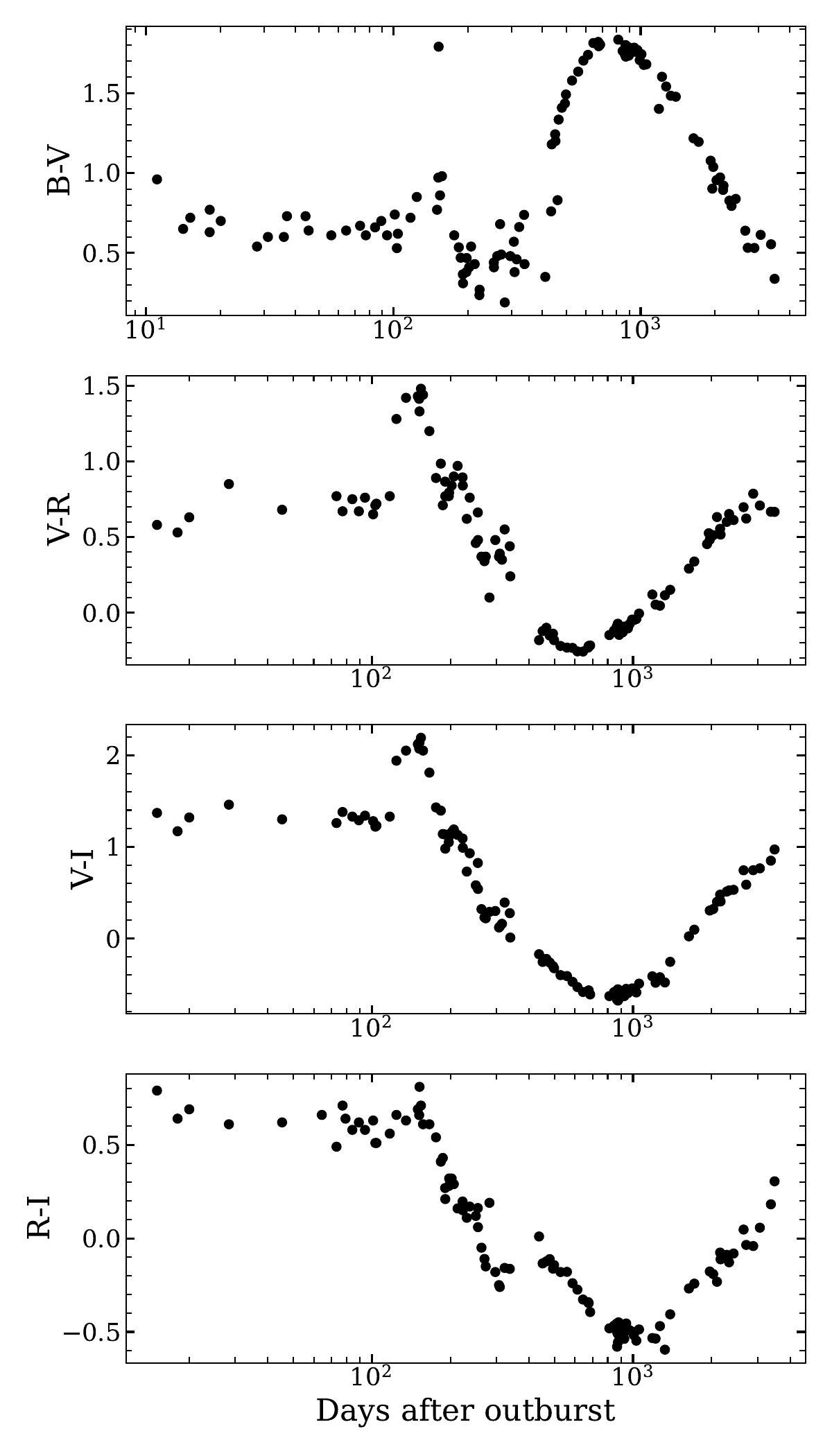}
\caption{Optical color of QY Mus from day 11 to 3482 generated using SMARTS and VSOJ data.}
	\label{op_color}
\end{figure}

The temporal evolution of the $B-V$, $V-R$, $V-I$, and $R-I$ color indices is shown in Fig.~\ref{op_color}. An increase in the optical colors is observed during the dust-dip phase. 
During the dust-dip phase (after $\sim$123 days), changes in the optical colors largely reflect variations in extinction caused by the formation and subsequent thinning of dust.
After day $\sim$200, the colors begin to evolve differently: $V-R$, $V-I$, and $R-I$ start to decline and become bluer, while $B-V$ continues to increase and becomes redder. This behaviour is consistent with the onset of the nebular phase (see Section~ \ref{Spectral evolution}), as evidenced by spectroscopic observations. 
During the nebular phase, strong emission lines in each filter dominate over the optical continuum in the $B$, $V$, $R$, and $I$ bands, which is reflected in the observed color evolution. As the [O\,\textsc{iii}] 4959, 5007~\AA~emission lines in the $V$ band increase in strength, the $B-V$ color becomes redder.
The H$\alpha$ and [N\,\textsc{ii}] 6545, 6584~\AA\ emission lines contributes significantly to the $R$ band. As H$\alpha$ fades relative to the [O\,\textsc{iii}] lines, the $V-R$ color decreases. The O\,\textsc{i} 8446 ~\AA~line contributes to the $I$ band, and although this line is not present in our nebular-phase spectra, the decreasing $R-I$ color is primarily due to the dominant contribution from the Balmer H$\alpha$ and [N\,\textsc{ii}] emission lines.
In the later stages, $B-V$ becomes bluer, while $V-R$, $V-I$, and $R-I$ become redder, likely indicating that the flux from nebular emission lines in the $V$ and $R$ bands has started to decline. After $\sim$2500 days, the colors do not change significantly, suggesting that the nova has entered into quiescence. This is further supported by its $B-V$ color in the CMD (Fig.~\ref{cmd_track}), which is in agreement with the region occupied by other quiescent novae \citep{2012ApJ...746...61D}.

\begin{figure}
\includegraphics[scale=0.33]{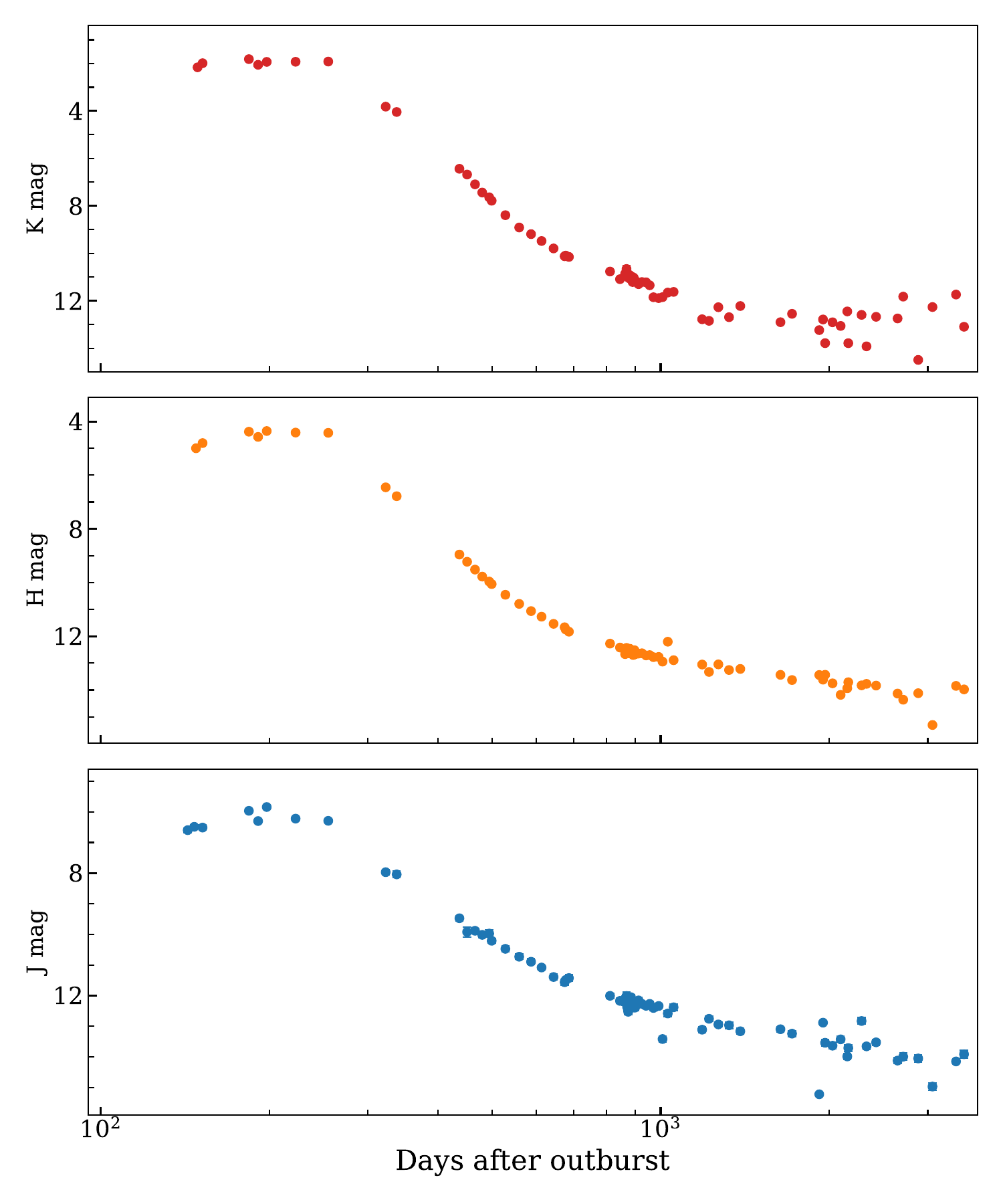}
\caption{NIR light curves of QY Mus from day 142 to 3482 generated using SMARTS.}
	\label{lc_nir}
\end{figure}

\subsection{NIR light curve}\label{nir_lc}

\begin{figure}
	\includegraphics[scale=0.40]{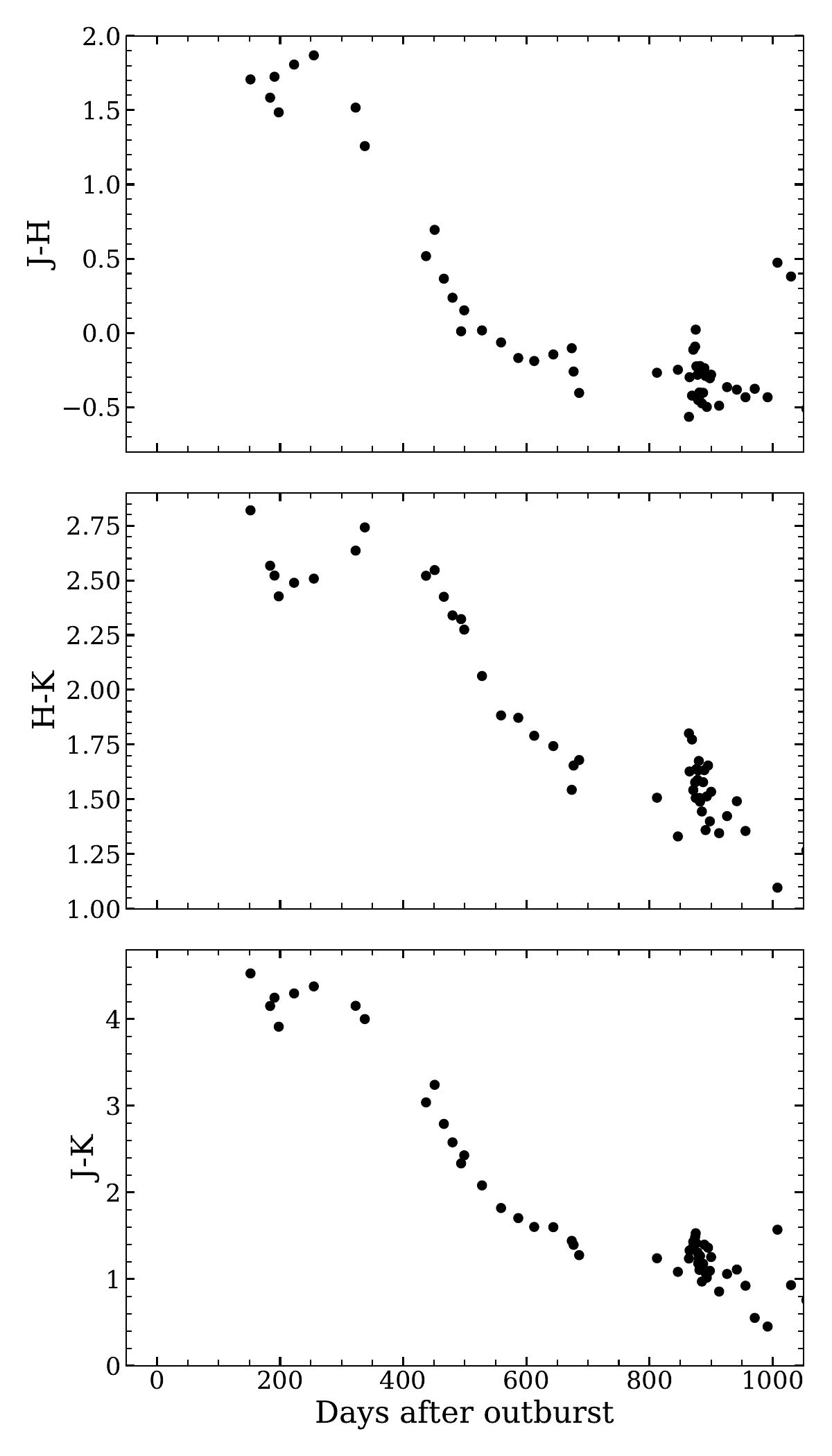}
\caption{NIR color of QY Mus from day 151 to 1050 generated using SMARTS.}
	\label{nir_color}
\end{figure}

\begin{figure}
    \centering
    \includegraphics[width=1.0\linewidth]{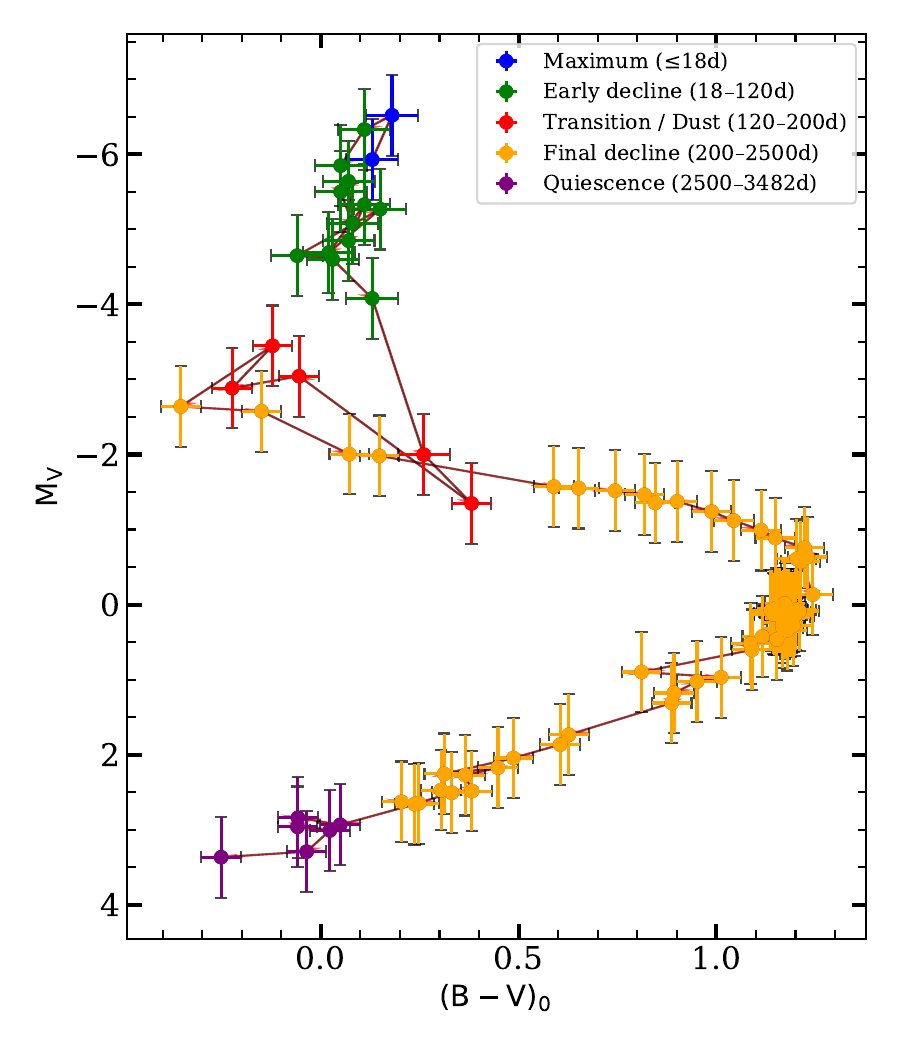}
    \caption{Color--magnitude diagram (CMD) of QY Mus constructed using $B$ and $V$ band photometry from VSOJ and SMARTS. The intrinsic color $(B-V)_0$ and absolute magnitude $M_V$ are derived after correcting for interstellar extinction. The distance is adopted from \citet{schaefer2022comprehensive}, estimated from Gaia parallax. The different colors represent various evolutionary phases.}
    \label{cmd_track}
\end{figure}

The NIR light curves, starting from 142 days since discovery and constructed using data from SMARTS, are presented in Fig.~\ref{lc_nir}. The NIR band magnitudes show a moderate increasing trend during the optical minimum phase, which is consistent with dust formation, since dust emission contributes mainly at longer wavelengths. 
The NIR color indices $J\!-\!H$, $H\!-\!K$, and $J\!-\!K$ exhibit an increase that is strongly affected by the presence of dust in the system, as shown in Fig.~\ref{nir_color}. The $J\!-\!K$, $J\!-\!H$, and $H\!-\!K$ indices attain peak values of 4.5, 1.9, and 2.8 magnitudes, respectively.
The brightness in the $JHK$ bands subsequently started declining, along with the color indices.  
During the dust condensation phase, thermal emission from dust increases the flux at longer wavelengths, then falls rapidly once grain growth stops. This behavior indicates that dust grains are either destroyed by radiation from the central source or that the dust shell is becoming more diffuse due to expansion \citep{gehrz2008infrared,shore2018spectroscopic}.

\subsection{Reddening and distance}\label{reddening and distance}
Based on average colors, consistent with previous findings, interstellar reddening can be estimated  using photometric data alone with an uncertainty of $\sim$0.2--0.3 mag.  \cite{1987A&AS...70..125V} reported that the novae at maximum brightness have an average intrinsic color of $(B - V)_0 = +0.23$ and $(B - V)_0 = -0.02$ at $t_2$.
Applying these values to our photometric data gives a reddening of 0.54 for the optical maximum and 0.64 at $t_2$.

More recently, \cite{10.1093/mnras/staf385}, based on their ``silver'' sample of novae, reported intrinsic colors of 
$(B - V)_0 = +0.20$ at optical maximum and $(B - V)_0 = -0.03$ at $t_2$. 
Applying these reference values to our data results in reddening estimates of 0.57 for optical maximum and  0.65 at $t_2$, 
both consistent with the values derived above.

The two-dimensional NASA/IPAC map of Galactic dust extinction provides an independent constraint of $E(B - V) = 0.58 \pm 0.03$ toward QY~Mus, which is consistent with that adopted in \cite{2019ApJS..242...18H}.
We therefore adopt a mean reddening of $E(B - V) = 0.59 \pm 0.05$. 
Taking $R_V = 3.1$, this corresponds to an interstellar extinction of $A_V = 1.84$.

Using the absolute magnitude at maximum (estimated in Section~\ref{optical light curve}) and the adopted reddening, we estimate a distance of $d = 4.64 \pm 1.16$ kpc from the distance modulus. 
The Gaia DR3 parallax for this source is $\varpi = 0.1115 \pm 0.1066$ mas. Given the considerable uncertainty in this measurement, direct inversion is not reliable. For comparison, \citet{2021AJ....161..147B} derived a distance of $d = 5.7^{+1.8}_{-2.4}$ kpc based on Gaia DR3 parallax. Independent work by \citet{schaefer2022comprehensive} reported a distance of $d = 4.57^{+1.48}_{-0.68}$ kpc based on Gaia parallax.  
The value given by \citet{schaefer2022comprehensive}, which is based on a Bayesian parallax calculation incorporating additional information from the nova peak magnitude, is more reliable and is in good agreement with our photometric distance. Our photometrically derived distance is fully consistent with both of these estimates. We adopt the distance estimate from \citet{schaefer2022comprehensive} for this work.

\subsection{Line identification, general characteristics, and evolution of the optical spectra}\label{Spectral evolution}
\begin{figure*}
\centering
	\includegraphics[scale=0.52]{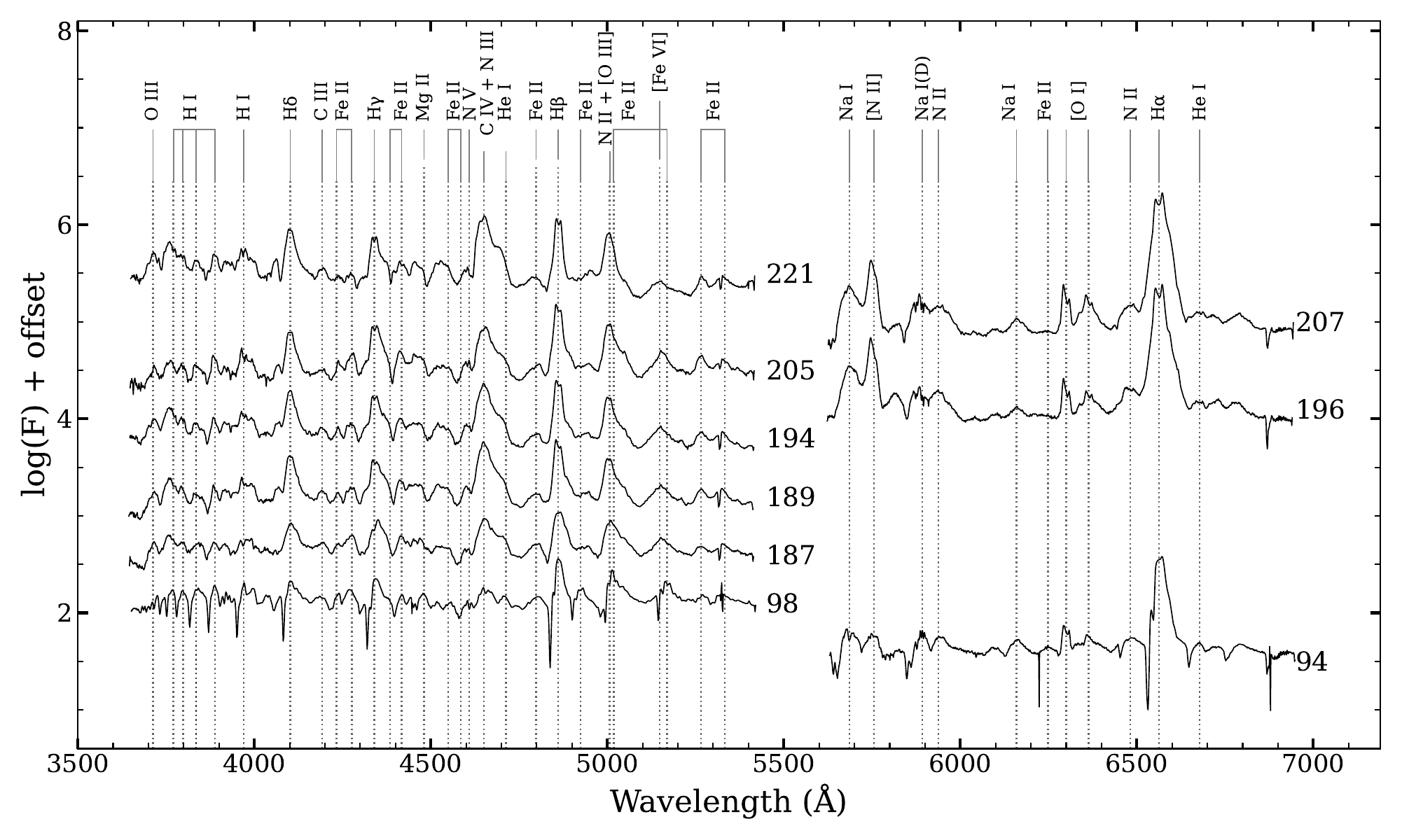}
\caption{
Spectroscopic evolution of QY~Mus during the early decline phase, from day 94 (2008 December 24) to day 221 (2009 May 1). The early spectra are dominated by Fe\,\textsc{ii} multiplets and hydrogen Balmer lines with P-Cygni profiles. The later spectra show a gradual transition to the nebular phase. Identified emission lines are labeled, and the time since discovery (in days) is indicated next to each spectrum.}
	\label{spectral1}
\end{figure*}

\begin{figure*}
\centering
    \includegraphics[scale=0.52]{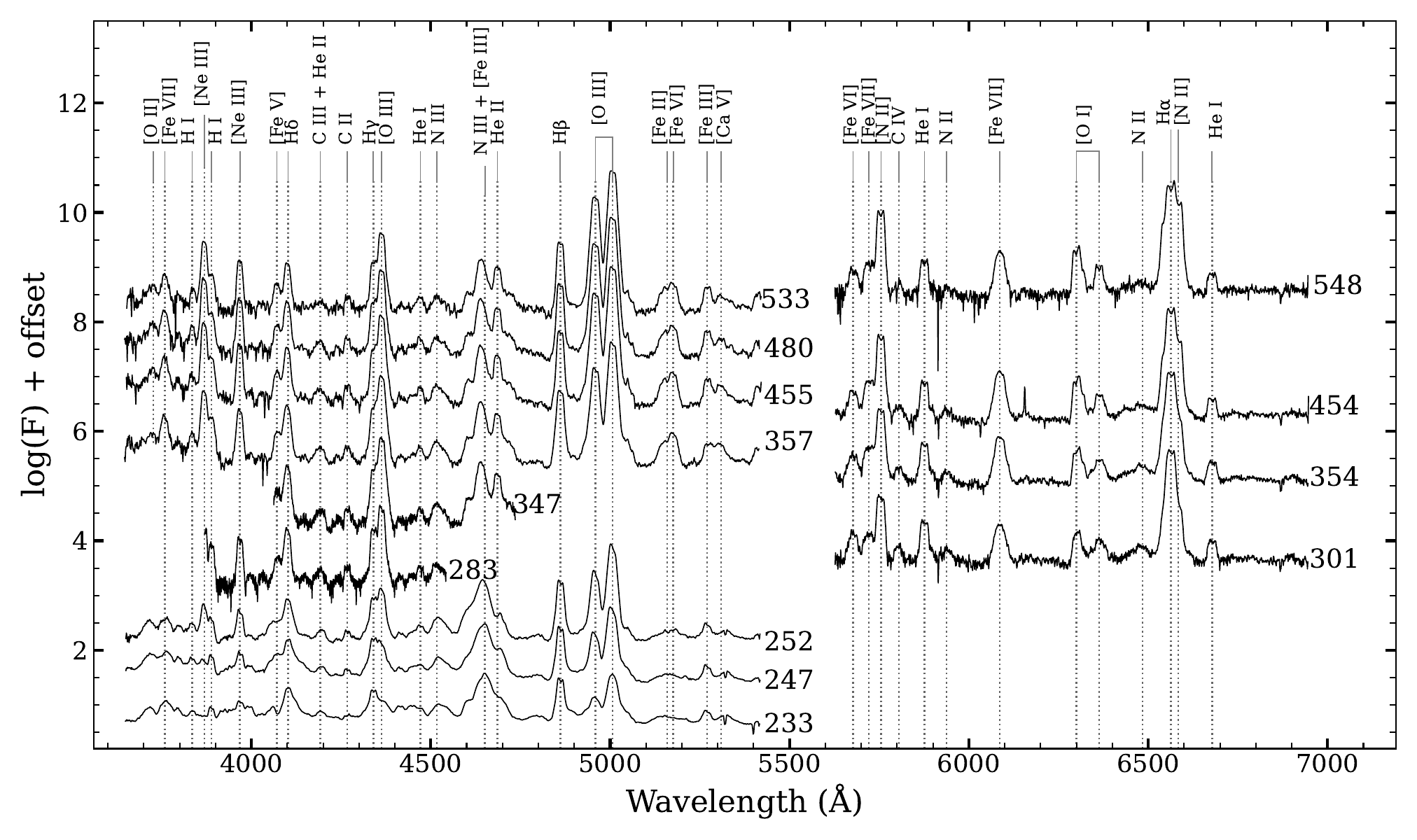}
\caption{During the onset of the nebular phase of Nova QY~Mus, the spectra obtained between day 233 (2009 May 13) and day 548 (2010 March 24) are dominated by strong forbidden emission lines of [O\,\textsc{iii}], [Ne\,\textsc{iii}], [Fe\,\textsc{vi}], and [Fe\,\textsc{vii}].}
	\label{spectral2}
\end{figure*}

\begin{figure*}
\centering
	\includegraphics[scale=0.52]{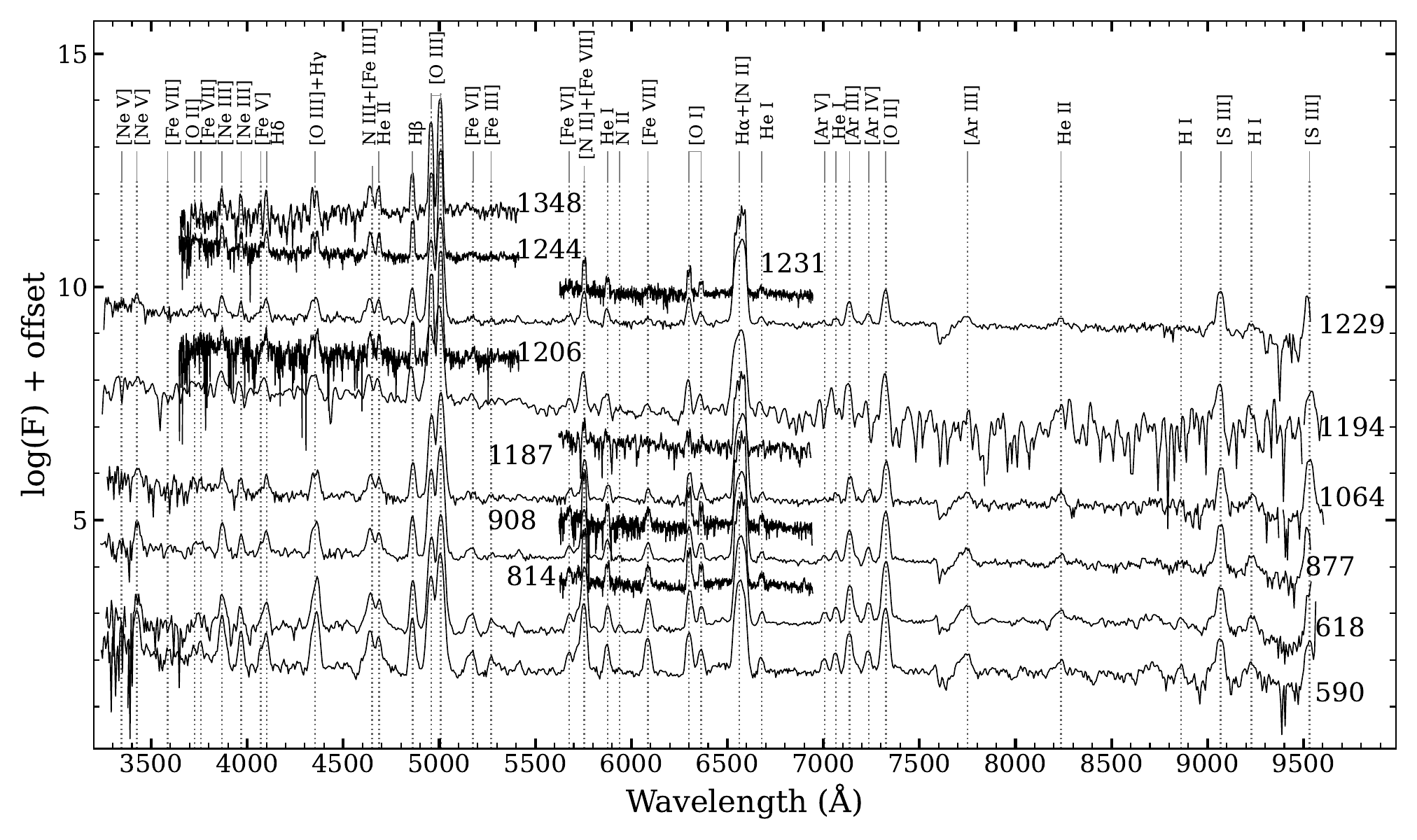}
\caption{Spectroscopic evolution of Nova QY Mus during the nebular phase, from day 590 (2010 May 05) to day 1348 (2012 June 01). The spectra display strong nebular phase lines of [O\,\textsc{iii}], [N\,\textsc{ii}], [Ne\,\textsc{v}], [Ne\,\textsc{iii}], [Ar\,\textsc{v}], [Ar\,\textsc{iv}], and [S\,\textsc{iii}].}
	\label{spectral3}
\end{figure*}

The spectroscopic evolution of QY~Mus, covering the early decline and nebular phases, is presented in Figs.~\ref{spectral1}, \ref{spectral2}, and \ref{spectral3}. 
Line identification was performed using the lines listed in \cite{williams2012origin} for novae. We present 35 low-resolution spectra spanning from 2008 December 23 to 2012 May 31 (days 94 to 1348 since outburst).

The early decline phase spectra obtained on days 94 and 98 show deep P-Cygni profiles in the Balmer lines (H$\alpha$, H$\beta$, H$\gamma$, and H$\delta$). 
Prominent Fe\,\textsc{ii} emission lines from multiplets 42, 48, 49, 57, and 58 are observed, with P-Cygni profiles notably at 4924, 5018, and 5169~\AA with absorption components at $-1450$~km~s$^{-1}$.
Other lines include [N\,\textsc{ii}] 5755~\AA, Na\,\textsc{i} 5686~\AA, Na\,\textsc{i} 6154/6160~\AA, and [O\,\textsc{i}] 6300 and 6364~\AA. The Na\,\textsc{i} D doublet at 5890 and 5896~\AA\ is also present, appearing as interstellar absorption features. Spectra from both epochs are consistent with the Fe\,\textsc{ii} class \citep{1992AJ....104..725W}. The prominent P-Cygni profiles observed on days 94 and 98 since outburst indicate slowly expanding ejecta.

Spectra obtained on days 187, 189, 194, and 196 show that the emission lines progressively grow stronger and more complex with time. The hydrogen Balmer lines and [O\,\textsc{i}] 6300, 6364~\AA\ later develop a double-peaked structure, indicating that the observed emission likely arises from a bipolar or equatorial ring structure of the ejecta \citep{hutchings1972non,2013ApJ...768...49R,2018ApJ...853...27M}. The full width at half maximum (FWHM) velocities of H$\beta$ and [O\,\textsc{i}] are measured to be 1580~km~s$^{-1}$ and 980~km~s$^{-1}$, respectively.
On days 205, 207, and 221, the [O\,\textsc{iii}] 4363, and 5007~\AA\ emission lines, along with the Bowen blend near 4640~\AA\ (primarily N\,\textsc{iii} and C\,\textsc{iii}), begin to appear in the spectra, while He\,\textsc{ii} 4686~\AA\ is visible as a distinct feature on the red side of the Bowen blend.
The 5007~\AA\ line is blended with  N\,\textsc{ii} 5001~\AA, and the 4363~\AA\ line is blended with H$\gamma$. The emergence of these [O\,\textsc{iii}] lines indicates that the nova is transitioning into the nebular phase. The spectra obtained from day 94 to 221 show lines of C\,\textsc{iii} 4187~\AA, N\,\textsc{v} 4609~\AA, C\,\textsc{iv} 4658~\AA, He\,\textsc{i} 4713~\AA, N\,\textsc{ii} 5001~\AA, N\,\textsc{ii} 5938~\AA, and N\,\textsc{ii} 6482~\AA, indicating a later He/N phase that appears before entering the nebular phase, as described in the new spectral classification by \citet{2024MNRAS.527.9303A}.

Spectra obtained on days 233, 247, 252, 283, and 301 (see Fig.~\ref{spectral2}) show increasingly strong nebular emission lines, including [O\,\textsc{iii}] 4363, 4959, and 5007~\AA, followed by the appearance of [N\,\textsc{ii}] 5755~\AA\ and the blend near 4640~\AA. Additional emission lines identified during this phase include [Fe\,\textsc{vii}] 3759~\AA, [Ne\,\textsc{iii}] 3869 and 3968~\AA, [Fe\,\textsc{v}] 4071~\AA\ blended with H$\delta$ 4102~\AA, C\,\textsc{iii} 4187~\AA\ / He\,\textsc{ii} 4200~\AA, He\,\textsc{i} 4471~\AA, [Fe\,\textsc{ii}] 5159~\AA, [Fe\,\textsc{vi}] 5176~\AA, [Fe\,\textsc{iii}] 5270~\AA, [Fe\,\textsc{vi}] 5677~\AA, [Fe\,\textsc{vii}] 5721~\AA\ blended with [N\,\textsc{ii}] 5755~\AA, He\,\textsc{i} 5876~\AA, [Fe\,\textsc{vii}] 6086~\AA, and He\,\textsc{i} 6678~\AA. The presence of strong [O\,\textsc{iii}], [Ne\,\textsc{iii}], and multiple ionization stages of iron ([Fe\,\textsc{ii}], [Fe\,\textsc{iii}], [Fe\,\textsc{v}], [Fe\,\textsc{vi}], and [Fe\,\textsc{vii}]) clearly indicates that the nova is in the nebular phase.
The optical colors also begin to reflect the dominance of emission lines over the optical continuum in the $B$, $V$, $R$, and $I$ bands, due to the presence of strong emission features within the respective filter passbands, resulting in the observed changes in color evolution (see Section~\ref{optical color} and Fig.~\ref{op_color}).
During the later epochs, from days 347 to 548, all of the above emission lines remain present, with [Ne\,\textsc{iii}] and [O\,\textsc{iii}] becoming increasingly prominent. The [O\,\textsc{iii}] 4363~\AA\ line strongly dominates the blend with the H$\gamma$ line. The 4640~\AA\ blend becomes resolved, with both N\,\textsc{iii} 4638~\AA\ and He\,\textsc{i} 4686~\AA\ clearly visible. H$\alpha$ is blended with [N\,\textsc{ii}] 6548 and 6584~\AA. The FWHM velocities of H$\beta$, [O\,\textsc{iii}] 5007~\AA, and [O\,\textsc{i}] 6300~\AA\ are measured to be 1100~km~s$^{-1}$, 1200~km~s$^{-1}$, and 1160~km~s$^{-1}$, respectively.

The full-range spectra obtained between days 590 and 1064 (see Fig.~\ref{spectral3}) show prominent emission lines of [Ne\,\textsc{v}] 3346 and 3426~\AA, [Ar\,\textsc{v}] 7006~\AA, [Ar\,\textsc{iii}] 7136 and 7751~\AA, [Ar\,\textsc{iv}] 7237~\AA, and [S\,\textsc{iii}] 9069 and 9561~\AA. From day 814 onward, the [N\,\textsc{ii}] 6584~\AA\ line begins to dominate the blend with H$\alpha$. The full spectrum obtained on day 1064 shows the strongest emission lines to be [O\,\textsc{iii}] 5007~\AA, followed by the blended H$\alpha$ and [N\,\textsc{ii}] lines, [O\,\textsc{iii}] 4959~\AA, [N\,\textsc{ii}] 5755~\AA, [O\,\textsc{ii}] 7320~\AA, and H$\beta$.
During the later epochs, from days 1187 to 1348, the emission lines of [Fe\,\textsc{vii}] 6086~\AA, [Fe\,\textsc{vi}] 5176~\AA, [Fe\,\textsc{vi}] 5677~\AA\, and [Fe\,\textsc{iii}] 5270~\AA~gradually weaken and disappear. In the final spectra obtained between days 1229 and 1348, the nova remains in the nebular phase, with strong forbidden emission lines of [Ne\,\textsc{v}], [Ne\,\textsc{iii}], [O\,\textsc{iii}], [N\,\textsc{ii}], [O\,\textsc{i}], [Ar\,\textsc{iii}], [Ar\,\textsc{iv}], [O\,\textsc{ii}], and [S\,\textsc{iii}]. The only permitted lines detected during this phase are H$\delta$, H$\beta$, and H$\alpha$, along with a few He\,\textsc{i} and He\,\textsc{ii} lines.

\begin{figure}
\centering
	\includegraphics[scale=0.39]{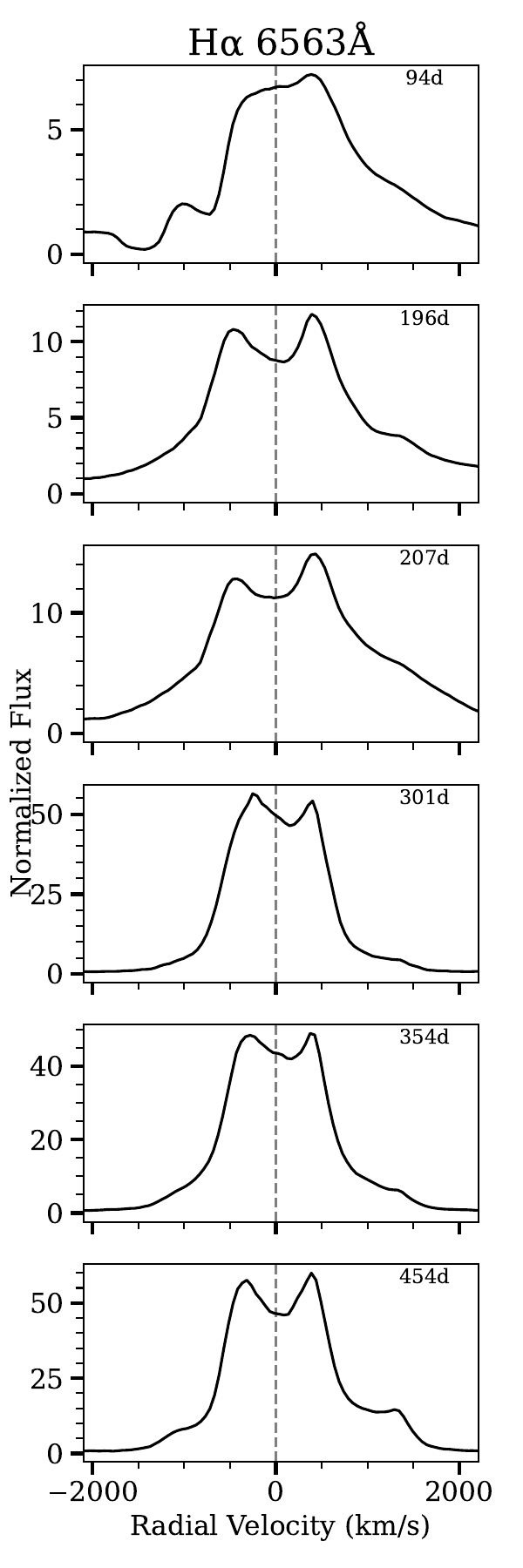}
    \includegraphics[scale=0.39]{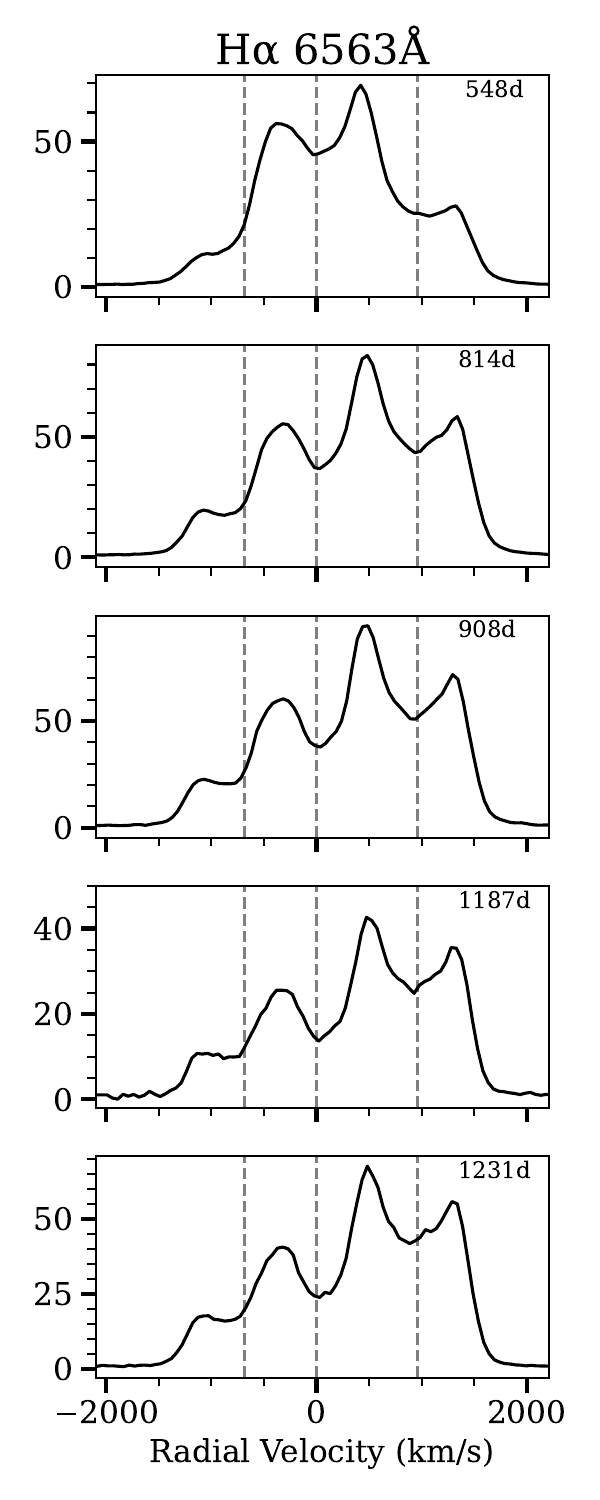}
\caption{Evolution of the H$\alpha$ velocity profile of QY Mus from day 94 to 1231 obtained from spectroscopic observations. The dashed vertical lines mark the rest wavelengths of H$\alpha$ and the [N\,\textsc{ii}] 6548, 6584~\AA~lines. The line profile evolves from a P-Cygni profile during the early decline phase to a double-peaked emission structure in the nebular phase (more details in section \ref{Spectral evolution}).}
	\label{h alpha}
\end{figure}

\begin{figure}
\centering
    \includegraphics[scale=0.39]{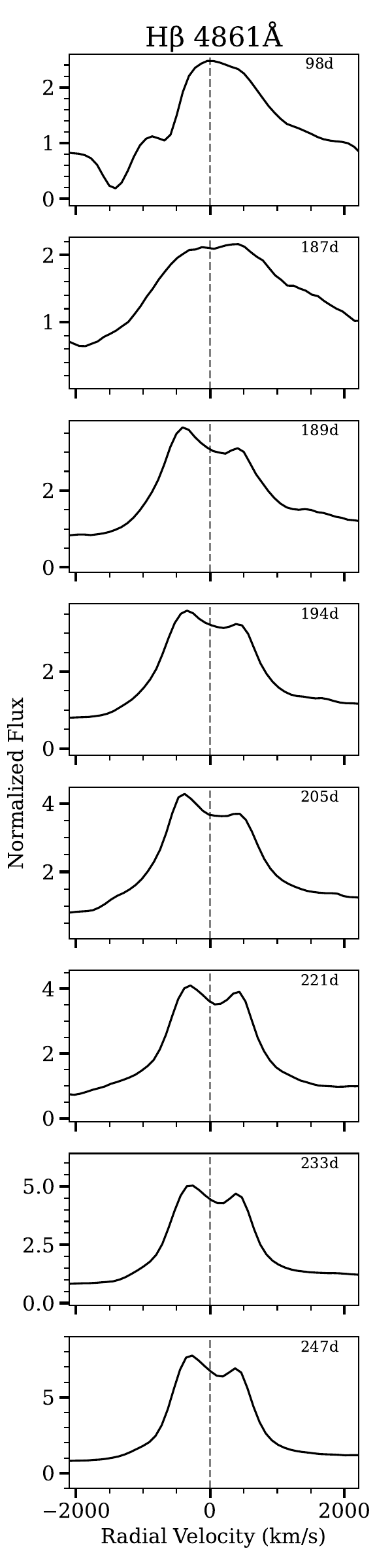}
    \includegraphics[scale=0.39]{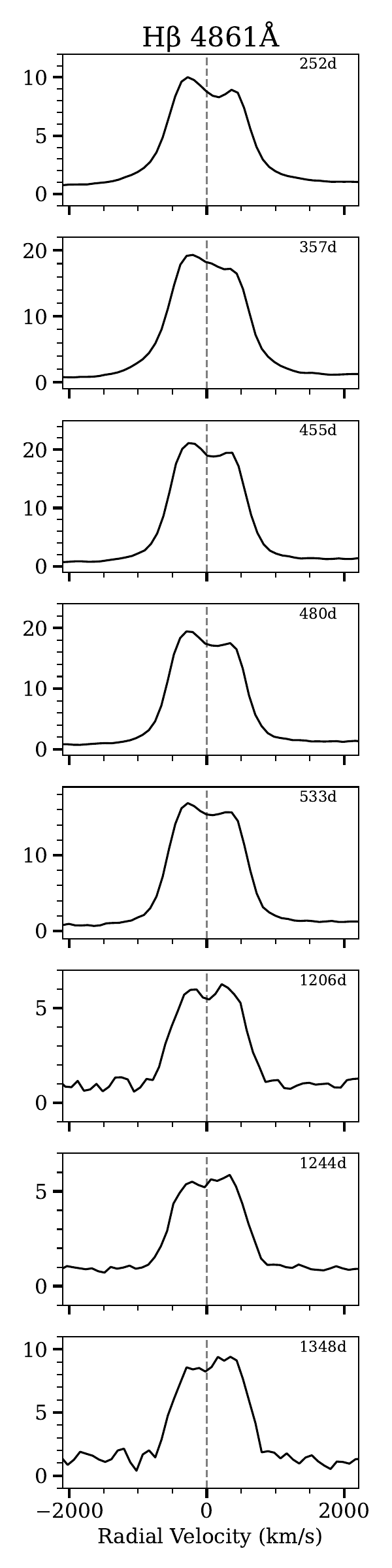}
\caption{The evolution of the H$\beta$ velocity profile of QY Mus from day 98 to day 1348 is shown. During the early decline phase, the line exhibits a P-Cygni profile, which transitions to a double-peaked emission structure in the nebular phase (more details in section \ref{Spectral evolution}.}
	\label{h beta}
\end{figure}

\begin{figure}
    \centering
    \includegraphics[scale=0.5]{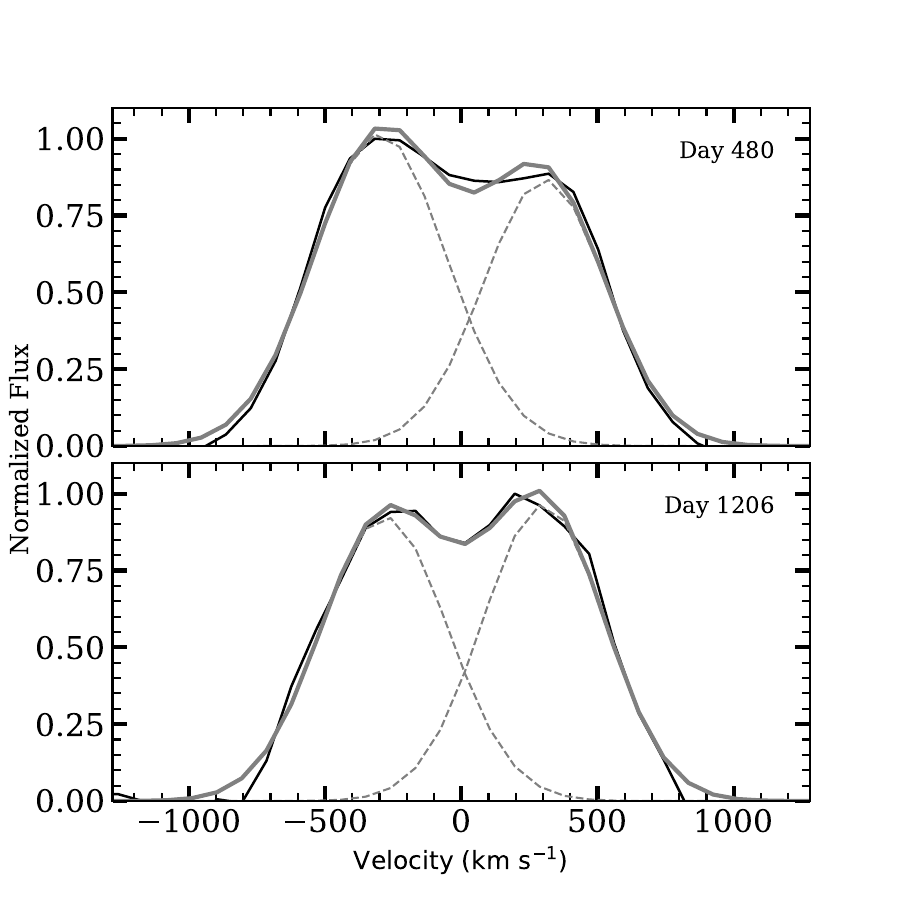}
    \caption{H$\beta$ line profiles on days 480 and 1206, showing a double-peaked structure. Dashed gray lines represent individual Gaussian components, the solid gray line shows their sum, and the solid black line indicates the observed data.}
    \label{gaussian}
\end{figure}

The evolution of the hydrogen Balmer H$\beta$ and H$\alpha$ line profiles shown in Figs.\ref{h alpha} and \ref{h beta} indicates that the early decline phase spectra exhibit prominent P-Cygni profiles. At later epochs, these profiles develop a double-peaked structure, likely indicative of a non-spherical ejecta geometry, commonly associated with a bipolar or ring-like structure of the ejecta \citep{2013ApJ...768...49R,2019A&A...622A.126P}. 
The H$\alpha$ line is blended with [N\,\textsc{ii}] 6548, and  6584~\AA~lines exhibit similar velocity profiles, resulting in a complex blended structure. In the later nebular phase, as the [N\,\textsc{ii}] emission becomes stronger, this blend evolves into a multi-peaked profile with up to four distinct components. It is seen that a two-component Gaussian fits the H$\beta$ line well, with the red- and blue-shifted components of the double-peaked profile having peaks at approximately the same velocity ($\sim$300~km~s$^{-1}$; see Fig.~\ref{gaussian}).

\subsection{Photoionization Modeling}\label{cloudy}
The full optical-range nebular-phase spectrum obtained on day 590 is used to derive the physical conditions and elemental abundances of the nova ejecta through photoionization modeling with the \textsc{Cloudy} code (version C23.01; \citealt{chatzikos20232023}). 
Among the available spectra, the complete optical-range spectrum obtained at day 590 is the most suitable for fitting photoionization models for the following reasons: it has the highest number of detectable lines (41), allowing the fit to cover a broad range of ionization stages. By day 590, the nova has completely evolved into its nebular phase, and the ejecta is optically thin, allowing photoionization models to be valid. Because the signal-to-noise ratio for this spectrum is much higher than those for the spectra from the later epochs, the continuum flux will be substantially lower in the later spectrum, resulting in a decrease in the number of detectable emission lines from those spectra, especially in the bluer end, will cause limitations in the multiple parameter \textsc{Cloudy} fit for those spectra. Furthermore, at later epochs, the strengthening of nebular emission leads to an increase in the blending of lines, which will reduce the accuracy in flux measurements. Therefore, we will restrict our fitting methodology to the day 590 spectrum for subsequent photoionization analysis.
\textsc{Cloudy} is a widely used photoionization code to model the physical conditions of non-equilibrium gas clouds exposed to an external radiation field. Using detailed microphysical processes, it self-consistently solves the equations of thermal and statistical equilibrium to compute the intensities and column densities of approximately $10^{4}$ spectral lines across the electromagnetic spectrum \citep{2017RMxAA..53..385F}. The resulting model-predicted line fluxes are directly compared with the observed spectra to constrain the physical properties of the source and the chemical composition of the gas cloud.

Our photoionization model assumes a spherically symmetric shell ionized by a central source with a blackbody spectrum characterized by a temperature $T_{\rm BB}$ (K) and luminosity $L$ (erg~s$^{-1}$). 
The shell is defined by parameters such as density, the radii of its inner and outer boundaries, the filling factor, and its elemental abundances.
In \textsc{Cloudy}, the density of the shell is defined by the total hydrogen number density $n(r)$ (cm$^{-3}$), given by
\begin{equation}
n(\mathrm{H}) = n(\mathrm{H}^{0}) + n(\mathrm{H}^{+}) + 2n(\mathrm{H}_{2}) + \sum_{\mathrm{other}} n(\mathrm{H}_{\mathrm{other}}),
\end{equation}
where $n(\mathrm{H}^{0})$, $n(\mathrm{H}^{+})$, $2n(\mathrm{H}_{2})$, and $n(\mathrm{H}_{\mathrm{other}})$ represent hydrogen in neutral, ionized, molecular, and all other hydrogen-bearing molecules, respectively.

The model considered abundances only for elements with detected emission lines in the spectra. For the other elements, we assumed solar abundances \citep{2010Ap&SSGrevesse}.
The elemental abundances in the model shell are defined using the abundance parameter in \textsc{Cloudy}. The expanding nova ejecta exhibit an inhomogeneous density structure, which includes dense clumps and lower-density diffuse regions \citep{paresce1995structure,BodeEvansBook2008,williams2013novae,chomiuk2021new}. To reproduce the observed line fluxes, a two-component density model is employed. This approach follows methodologies established in previous photoionization modeling \citep[e.g.,][]{shore2003early,2010AJHelton,raj2018cloudy2676,2020MNRASPavana,raj2024dustyaftermathrapidnova}.
The modeling methodology followed in this study is described in detail in \citet{raj2018cloudy2676,raj2024dustyaftermathrapidnova,Bisht_2025,bisht2025spectrophotometric}.

\begin{figure*}
    \centering
    \includegraphics[width=0.90\linewidth]{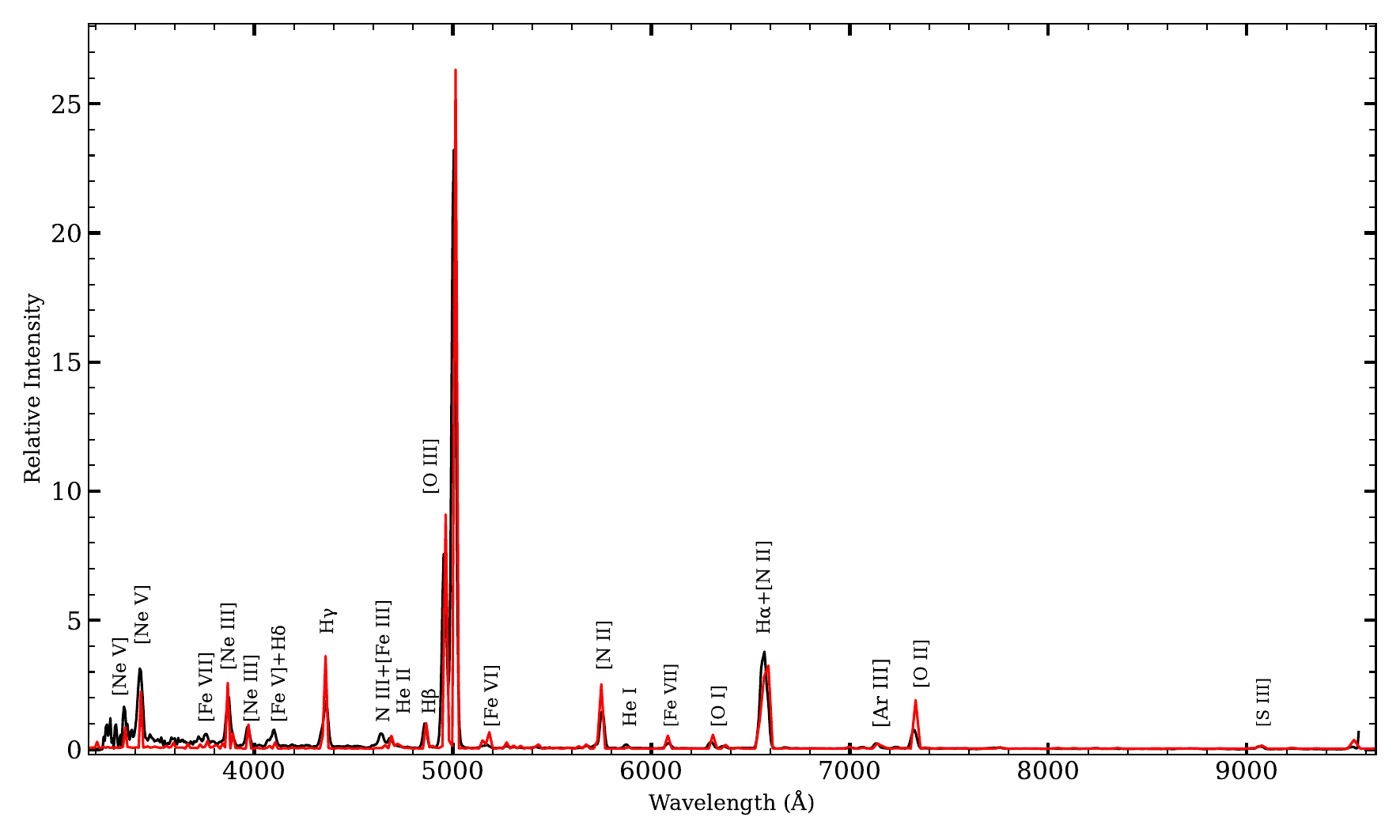}
    \caption{Comparison between the observed optical spectrum of QY~Mus (day 590, black) and the synthetic spectrum derived from the optimized \textsc{Cloudy} model (red).}
    \label{cloudy79}
\end{figure*}

The $E(B-V)$ value estimated in Section~\ref{reddening and distance} was used to deredden the observed emission lines. Fluxes from 41 lines in the day 590 spectrum were then employed for model fitting.
The best-fit model is determined from the $\chi^{2}$ and reduced $\chi^{2}$ tests. The dereddened observed and model relative fluxes are presented in Table~\ref{d590_flux_table}, along with their corresponding $\chi^{2}$ values. 
Table~\ref{d509_phy_opt} presents the best-fit physical parameters and elemental abundances obtained from the model. Figure~\ref{cloudy79} compares the observed spectrum with the synthetic spectrum generated from the model parameters.

By day 590, the spectrum was dominated by high-ionization forbidden lines, indicating that the system had entered the nebular phase. The modeling results estimate that the temperature of the ionizing source at this stage is $(7.08 \pm 0.20)\times10^{5}$~K.
The derived central source temperature is physically consistent with the expected behavior of the white dwarf at this evolutionary stage. By day 590, the nova had been in the nebular phase, with the white dwarf envelope largely depleted and residual hydrogen burning expected to be in its late decline. Theoretical models of post-outburst white dwarf cooling (e.g., \cite{sala2005envelope,wolf_2013}) predict effective temperatures within this range, supporting the derived value. Additionally, at this temperature, high-ionization lines such as [Ne\,\textsc{v}], [Fe\,\textsc{vii}], [Fe\,\textsc{vi}], and [Fe\,\textsc{v}] are expected to arise, providing qualitative validation of our modeling result.
The higher-ionization lines [Fe\,\textsc{vii}], [Fe\,\textsc{vi}], [Ne\,\textsc{v}], and [Ar\,\textsc{v}] are produced mainly in regions of lower density, whereas lower-ionization lines such as [S\,\textsc{iii}], [Ne\,\textsc{iii}], [Ar\,\textsc{iii}], [Fe\,\textsc{iii}], [N\,\textsc{ii}], [O\,\textsc{i}], and He\,\textsc{i} originate from relatively higher-density regions. The H\,\textsc{i}, [O\,\textsc{iii}], and He\,\textsc{ii} emission lines arise from both the high- and low-density components of the ejecta.
Our abundance results from the photoionization modeling indicate that nitrogen, oxygen, and neon are enriched relative to their solar values. The enhanced nitrogen abundance in the ejecta is explained by proton-capture during the thermonuclear runaway (TNR) on the surface of the white dwarf \citep{Starrfield_2016}. Enrichments of this kind are commonly reported in classical novae and are understood to arise from a combination of nuclear processing during the outburst and mixing between the accreted envelope and the underlying white dwarf material \citep{1994AJ....108.1008H,1999PhR...311..405G,2018A&A...619A.121C,Starrfield2020}.

The neon emission lines [Ne V] and [Ne III] in QY~Mus are stronger but weak with respect to the [O III] 5007~\AA, line, and our model reproduces them with a modest neon abundance of Ne/H = $2.2 \pm 0.5$. In contrast, the neon novae such as Nova LMC~1990 No.~1, QU~Vul, and V5583~Sgr exhibited strong neon lines stronger than the [O III] 5007~\AA~ line, indicating substantial neon enrichment \citep{1991ApJ...376..721W,2002ApJ...577..940S,2014MNRAS.438.3483H}. 
A defining characteristic of ONe novae is a significant abundance of neon, typically exceeding $\sim20$ times the solar value \citep{1994ApJ...425..797L, Schwarz_2007,2012ApJ...755...37H}.
The significantly lower abundance of neon compared to the ONe novae derived for QY~Mus therefore suggests that the nova QY~Mus does not belong to the neon nova class.

The results of the best-fit model parameters were used to calculate the hydrogen mass within the modeled shell using the equation given by \citet{2001MNRAS.320..103S}:
\begin{equation}
M_{\text{shell}} = n(r_{\text{in}})\, f(r_{\text{in}}) 
\int_{r_{\text{in}}}^{r_{\text{out}}}
\left(\frac{r}{r_{\text{in}}}\right)^{\alpha+\beta}
4\pi r^{2}\, dr .
\end{equation}

Using the model parameters, the mass of the ejected hydrogen shell is estimated to be $(1.34 \pm 0.08) \times 10^{-4}\, M_{\odot}$ on day 590.

The photoionization model presented in this study assumes a spherical-shell geometry. However, the spectroscopic evolution of QY Mus, in particular the double-peaked profiles of the Balmer and [O i] lines and the multi-component structure of the H$\alpha +$[N\,\textsc{ii}] blend (Sections \ref{Spectral evolution}, Figs. \ref{h alpha} - \ref{h beta}), provides clear evidence for a non-spherical ejecta geometry, most likely bipolar or equatorial ring-like. The assumption of spherical symmetry is a standard simplification widely adopted in nova photoionization studies (e.g., \cite{2010AJHelton,raj2018cloudy2676,2020MNRASPavana}) and is motivated here by the fact that Cloudy operates in 1D and does not natively support full 3D geometric configurations.

The primary consequences of this simplification are as follows. The filling factor and covering factor parameters in the model implicitly account for some geometric complexity, partially compensating for the assumed symmetry. Elemental abundances derived from emission line ratios remain relatively robust to geometric assumptions, as these ratios are predominantly sensitive to temperature and ionization structure rather than the precise spatial distribution of gas. Studies of other novae with asymmetric ejecta morphologies, such as bipolar or equatorial ring structures inferred from morpho-kinematic analyses (e.g., \cite{shore2003early,2020MNRASPavana,basu2024}), have demonstrated that spherically symmetric Cloudy models reproduce observed line fluxes to within approximately 20–30 $\%$, with abundances typically consistent within the quoted uncertainties.

To further evaluate the sensitivity of the results to the geometric assumption, a cylindrical geometry was also tested following an approach similar to \citep{wood_2024}. In Cloudy, a cylindrical geometry is simulated using a truncated sphere \citep{ferland_1982}. All best-fit parameters from the spherical model (Table \ref{d509_phy_opt}) were held constant, with only the semi-height of the cylinder allowed to vary. The best-fitting cylindrical model yields a semi-height of $h_{\rm cyl} = 7.13\times10^{15}$ cm. This cylindrical model produces emission-line fluxes comparable to those of the spherical case, quantitatively demonstrating that the derived physical conditions, including central source temperature, hydrogen densities, and elemental abundances, are not strongly sensitive to the assumed geometry. Therefore, the spherical model provides a reliable characterization of the physical conditions in the nova ejecta, and the parameters reported in Table \ref{d509_phy_opt} are robust within the quoted uncertainties.

\begin{table}
	\caption{Emission-line flux ratios for QY~Mus on day 590, presenting both the observed values and those predicted by the best-fitting \textsc{Cloudy} model, along with the corresponding $\chi^2$ value.}
	\label{d590_flux_table}
	\begin{center}
		\resizebox{\hsize}{!}{%
			\begin{tabular}{lclcc}
				\hline
				\hline
				\textbf{Line ID} & \boldmath{$\lambda$ (\AA)} & \textbf{Observed}$^{a}$ & \textbf{Modelled}$^{a}$ & \boldmath{$\chi^2$}  \\
				\hline
			    {[}Ne V{]}       & 3346       & 1.25E$+$00 & 1.07E$+$00 & 3.54E$-$01 \\
				{[}Ne V{]}    & 3426       & 3.65E$+$00 & 2.94E$+$00 & 5.80E$+$00 \\
				  {[}Fe VII{]}   & 3586       & 2.24E$-$01 & 2.63E$-$01 & 1.72E$-$02 \\
				{[}S III{]} + {[}Fe III{]}     & 3720    & 3.70E$-$01 & 1.22E$-$01 & 6.80E$-$01 \\
                {[}Fe VII{]}   & 3759       & 4.09E$-$01 & 3.76E$-$01 & 1.725E$-$02 \\
                {[}Ne III{]}       & 3869       & 2.06E$+$00 & 2.29E$+$00 & 5.71E$-$01 \\
                {[}Fe V{]}    & 3891       & 1.37E$-$01 & 2.68E$-$01 & 1.925E$-$01 \\
                {[}Ne III{]}       & 3968       & 8.00E$-$01 & 6.95E$-$01 & 1.74E$-$01 \\
                {[}Fe V{]}    & 4071       & 1.43E$-$01 & 1.01E$-$01 & 1.775E$-$02 \\
                H I           & 4102       & 6.86E$-$01 & 2.48E$-$01 & 1.57E$+$00 \\
                H I           & 4340       & 5.72E$-$01 & 4.76E$-$01 & 7.72E$-$02 \\
                {[}O III{]}   & 4363       & 2.00E$+$00 & 3.57E$+$00 & 20.22E$+$00 \\
				  N III + {[}Fe III{]} & 4640& 6.70E$-$01 & 1.01E$-$01 & 3.59E$+$00 \\
				He II         & 4686       & 3.49E$-$01 & 5.72E$-$01 & 5.55E$-$01 \\
				H I           & 4861       & 1.00E$+$00 & 1.00E$+$00 & 0.00E$+$00 \\
				  {[}O III{]}   & 4959       & 9.56E$+$00 & 9.49E$+$00 & 6.40E$-$02 \\
				  {[}O III{]}   & 5007       & 28.81E$+$00 & 28.32E$+$00 & 2.68E$+$00 \\
                {[}Fe VI{]}    & 5176       & 2.38E$-$01 & 7.71E$-$01 & 3.14E$+$00 \\
				{[}Fe III{]}    & 5270       & 1.11E$-$01 & 5.68E$-$02 & 3.38E$-$02 \\
				 {[}Fe VI{]}    & 5677       & 1.24E$-$01 & 1.96E$-$01 & 5.94E$-$02 \\
                 {[}Fe VII{]}    & 5721       & 1.20E$-$01 & 4.26E$-$01 & 7.64E$-$01 \\
				{[}N II{]}    & 5755       & 1.79E$+$00 & 1.83E$+$00 & 1.52E$-$02 \\
				He I          & 5876       & 1.76E$-$01 & 4.33E$-$02 & 1.96E$-$01 \\
				{[}Fe VII{]}    & 6086       & 3.15E$-$01 & 6.37E$-$01 & 1.15E$+$00 \\
                {[}O I{]}     & 6300       & 3.70$-$01 & 3.01E$-$01 & 7.78E$-$02 \\
				{[}O I{]}     & 6364       & 1.34E$-$01 & 1.00E$-$01 & 2.30E$-$02 \\
				{[}N II{]}    & 6548       & 9.46E$-$01 & 7.64E$-$01 & 2.68E$-$01 \\
                H I           & 6563       & 3.98E$+$00 & 2.85E$+$00 & 10.43E$+$00 \\
				{[}N II{]}    & 6584       & 2.39E$+$00 & 2.25E$+$00 & 1.62E$-$01 \\
                He I          & 6678       & 5.62E$-$02 & 8.60E$-$03 & 2.25E$-$02 \\
                {[}Ar V{]}    & 7006       & 4.78E$-$02 & 8.20E$-$02 & 1.29E$-$02 \\
                He I          & 7065       & 7.72E$-$02 & 2.45E$-$02 & 4.42E$-$02 \\
                {[}Ar III{]}  & 7136       & 3.11E$-$01 & 1.72E$-$01 & 2.18E$-$01 \\
				{[}Ar IV{]}  & 7237       & 9.29E$-$02 & 8.87E$-$03 & 7.85E$-$02 \\
                {[}O II{]}    & 7325       & 1.11E$+$00 & 8.05E$-$01 & 1.09E$+$00 \\
                {[}Ar III{]}  & 7751       & 1.66E$-$01 & 4.07E$-$02 & 1.75E$-$01 \\
                 He II          & 8237       & 4.96E$-$02 & 1.30E$-$02 & 1.50E$-$02 \\
                 H I           & 8665       & 3.03E$-$02 & 1.33E$-$02 & 3.18E$-$03 \\
                {[}S III{]}  & 9069       & 2.19E$-$01 & 1.01E$-$01 & 1.64E$-$01 \\
                 H I           & 9229       & 4.21E$-$02 & 2.24E$-$02 & 4.32E$-$03 \\
                {[}S III{]}  & 9531       & 1.03E$-$01 & 2.44E$-$01 & 2.21E$-$01 \\
				\hline
		    \end{tabular}}
	\end{center}
	$^{a}$Relative to H$\beta$
\end{table}

\begin{table}
	\caption{Best-fit optical \textsc{Cloudy} model parameters obtained on day 590 for the system QY Mus}
	\label{d509_phy_opt}
	\begin{center}
		\resizebox{\hsize}{!}{%
			\begin{tabular}{l c } 
				\hline\hline
				\textbf{Parameter} & \textbf{Day 590} \\ [0.5ex] 
				\hline 
				T$_{BB}$ ($\times$ 10$^{5}$ K) & 7.08 $\pm$ 0.20 \\ [0.25ex]
				Luminosity ($\times$ 10$^{37}$ erg/s) & 1.00 $\pm$ 0.25 \\ [0.25ex]
				Clump Hydrogen density ($\times$ 10$^{6}$ cm$^{-3}$) & 2.30 $\pm$ 0.03 \\ [0.25ex]
				Diffuse Hydrogen density ($\times$ 10$^{6}$ cm$^{-3}$) & 1.51 $\pm$ 0.03\\ [0.25ex]
				Covering factor (clump) & 0.30 \\ [0.25ex]
				Covering factor (diffuse) & 0.70 \\ [0.25ex]
				$\alpha$ & -3.00 \\ [0.25ex]
				Inner radius ($\times$ 10$^{15}$ cm) & 5.96 \\ [0.25ex]
				Outer radius ($\times$ 10$^{15}$ cm) & 8.41 \\ [0.25ex]
				Filling factor  & 0.10 \\ [0.25ex]
				N/N$_{\odot}$ & 39.00 $\pm$ 8.0 (3)$^{a}$ \\ [0.25ex]
				O/O$_{\odot}$ & 3.70 $\pm$ 0.8 (5) \\ [0.25ex]
				Ne/Ne$_{\odot}$ & 2.20 $\pm$ 0.50 (4) \\ [0.25ex]
				Ejected mass ($\times$ 10$^{-4}$ M$_{\odot}$) & 1.34 $\pm$ 0.08\\ [0.25ex]
				Number of observed lines (n) & 41 \\ [0.25ex]
				Number of free parameters (n$_{p}$) & 9 \\ [0.25ex]
				Degrees of freedom ($\nu$) & 32 \\ [0.25ex]
				Total $\chi^{2}$ & 55.13 \\ [0.25ex]
				$\chi^{2}_{red}$ & 1.72 \\ [0.25ex]
				\hline
		\end{tabular}}
	\end{center}
	$^{a}$The number of lines availed to obtain abundance estimate is as shown in the parenthesis.
\end{table}

\subsection{Dust mass and temperature}\label{dust masss and temp}
\begin{figure}
\centering
	\includegraphics[scale=0.30]{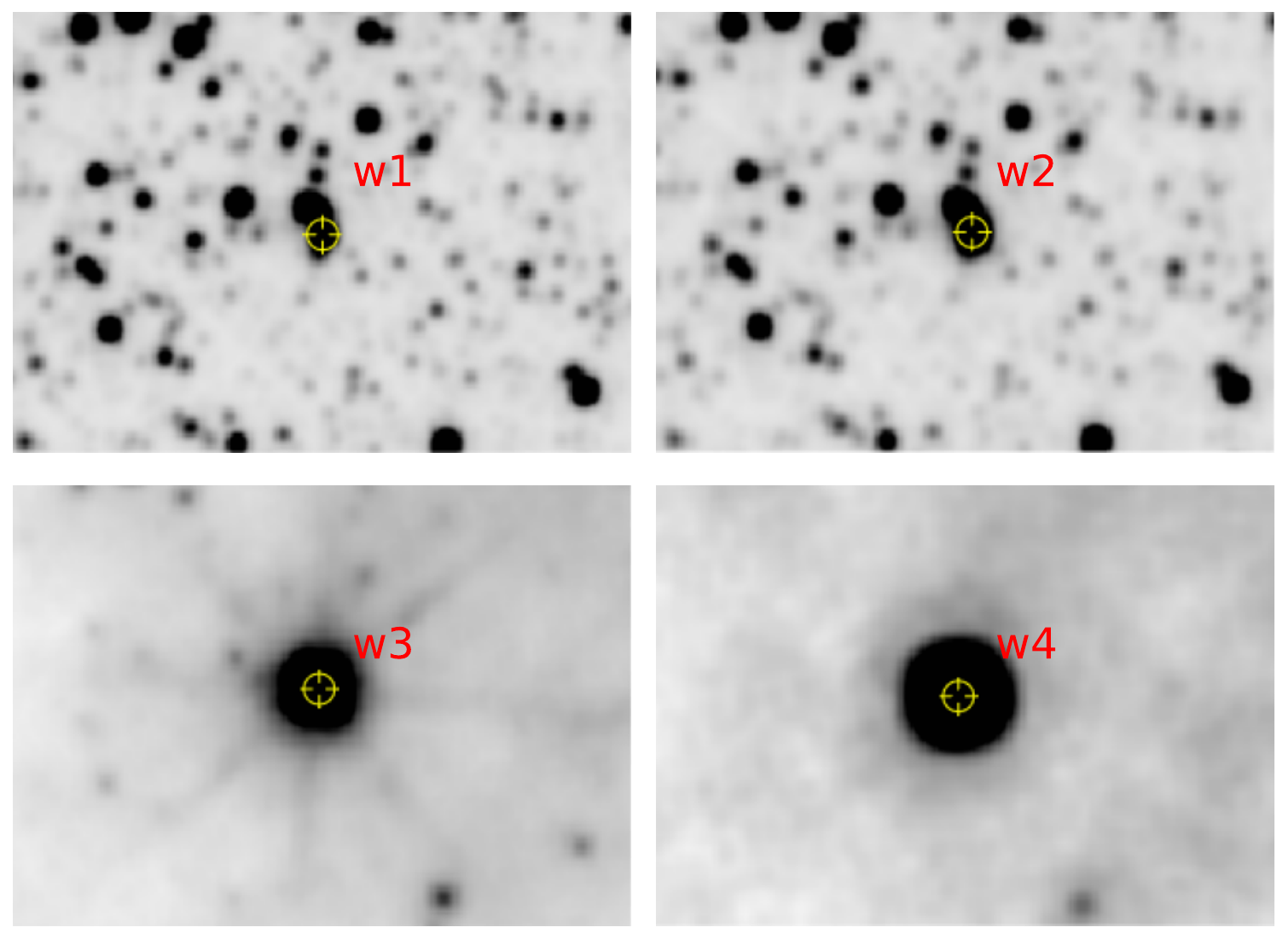}
	\caption{Image from the WISE survey, using combined data from observations carried out between February 2010 and August 2010. The 3$\times$3 arcmin field around QY~Mus shows clear detection in all four mid-infrared bands, with particularly strong emission in the W3 and W4 bands (see Section~\ref{dust masss and temp}).}
	\label{wise_image}
\end{figure}

The sudden decline in the optical light curve accompanied by a simultaneous rise in the near-infrared emission provides clear evidence for dust formation in the ejecta, classifying the nova as a dust-forming D-type. Mid-infrared observations from the WISE in the W1 (3.4~$\mu$m), W2 (4.6~$\mu$m), W3 (12~$\mu$m), and W4 (22~$\mu$m) bands reveal significant emission at longer wavelengths by dust, on $\sim$509 days since discovery (see Fig.~\ref{wise_image}). The emission is particularly prominent in the W3 and W4 bands, indicating the presence of relatively cool dust.

\begin{figure}
    \centering
    \includegraphics[width=1.0\linewidth]{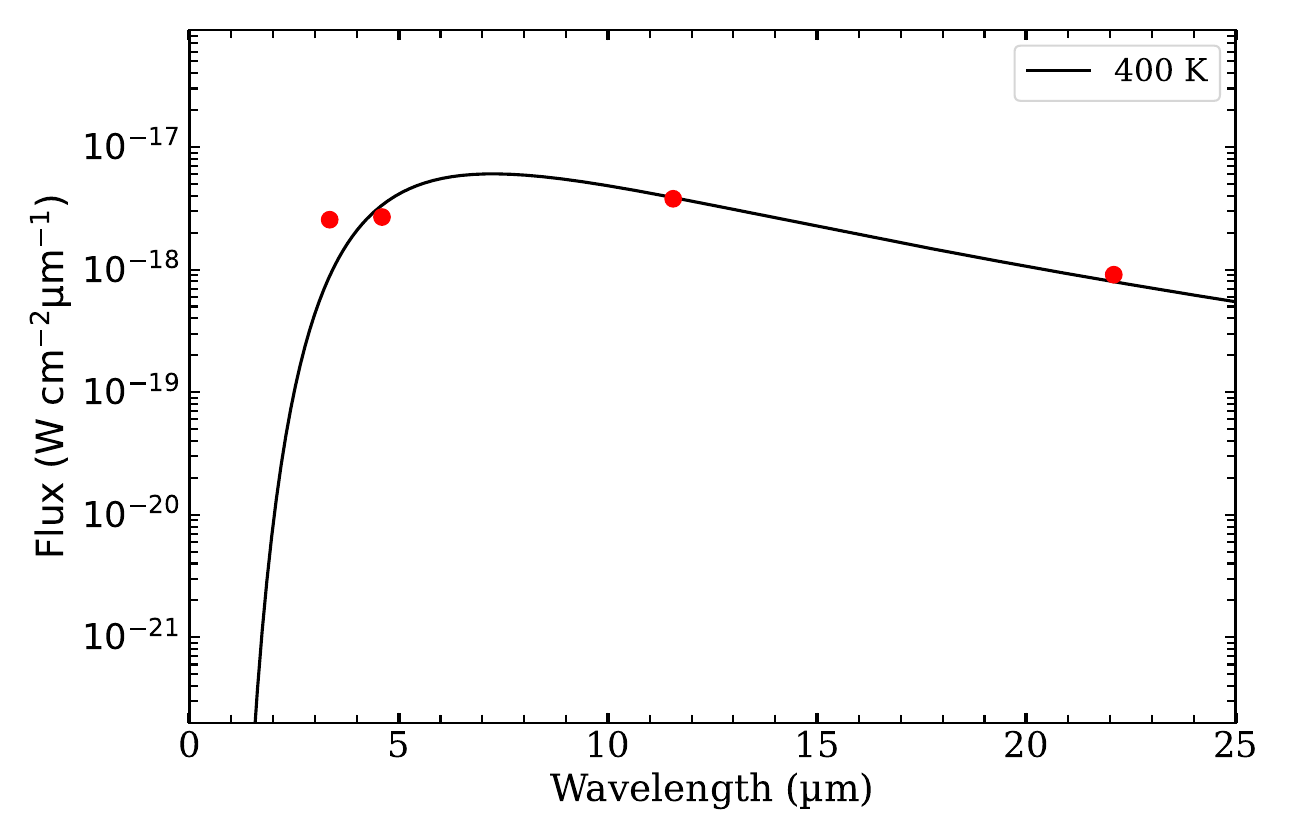}
    \caption{The SED shows a best fit blackbody to the WISE data taken on day 509, with a temperature of about 400
K.}
    \label{sed}
\end{figure}

Using the WISE photometry, we estimated the dust temperature at $\sim$509 days to be   400 $\pm$ 50 K. This estimate is subject to considerable uncertainty, as it assumes isothermal dust and is based on only four mid-infrared data points. The spectral energy distribution peaks near the W3 band ($\sim$8.3~$\mu$m; Fig.~\ref{sed}), consistent with thermal emission from cool dust. 

We estimated the dust mass following the relations by \citet{evans2017rise}, assuming spherical grains composed of carbonaceous material. Adopting the distance from section \ref{reddening and distance}, a grain density of $\rho = 2.25~{\rm g~cm^{-3}}$, and using $(\lambda F_\lambda)_{\rm max}$ in units of W~m$^{-2}$, we derive dust masses of 
$M_{dust,AC} = (8.11 \pm 4.28) \times 10^{-8}\,M_\odot$ for amorphous carbon grains and 
$M_{dust,GR}= (2.51 \pm 1.32) \times 10^{-7}\,M_\odot$ for graphitic carbon grains.

The long-term fate of the dust formed in QY Mus ejecta can be inferred from the photometric evolution. Following the dust minimum around day 146, the optical light curve began to recover by day $\sim$200, indicating a reduction in the effective optical depth of the dust shell. This recovery is consistent with two non-mutually exclusive processes: (i) grain destruction by the increasing UV and X-ray flux from the still active white dwarf, which can photodissociate carbonaceous grains on timescales of months to years post-outburst \citep{gehrz2008infrared}; and (ii) geometric dilution, whereby the expanding ejecta causes the dust shell to thin and become progressively more optically transparent even if the total grain mass remains approximately constant. The NIR color indices (J-K, H-K; Fig. \ref{nir_color}) begin to decline after their peak values, consistent with cooling and dispersal of the dust. The WISE mid-IR observations at $\sim$509 days confirm that cool dust ( $\sim400$ K) persisted well beyond the optical recovery, suggesting that complete grain destruction had not occurred by this epoch. The dust mass estimates of (8.11 $\pm 4.28) \times 10^{-8}\,M_\odot$ (amorphous carbon) and (2.51 $\pm 1.32) \times 10^{-7}\,M_\odot$ (graphitic carbon) represent a snapshot at this epoch. At later epochs, no mid-IR data are available to track the further evolution of the dust; however, by analogy with other dust-forming novae such as V5579 Sgr \citep{raj2024dustyaftermathrapidnova} and V1280 Sco \citep{sakon2016concurrent}, the dust is expected to have continued cooling and dispersing as the ejecta expanded, eventually becoming undetectable at mid-infrared wavelengths on timescales of a few years post-outburst.

\section{Discussion}\label{Discussion}
From the dip at $\sim$123 days in the optical light curve and the subsequent recovery around 200 days, it is evident that QY~Mus formed dust in the ejecta. During this phase, the NIR colors become redder, with $J\!-\!K \sim 4.5$ mag. In comparison with other dust-forming novae, similarly high color indices have been reported, including 3.8 for V496~Sct \citep{raj2012v496}, 8.0 for V2676~Oph \citep{raj2017v2676}, 6.7 for V5579~Sgr \citep{raj2024dustyaftermathrapidnova}, and 5.6 for LMCN~2009-05a \citep{Bisht_2025}.
The polynomial fit in the $V$ band shows that the $V$ band declines by $\tau_{\rm max} \sim 3.2$ mag within 23 days after the onset of dust, due to obscuration of the optical pseudo-photosphere by dust \citep{gehrz1988infrared}. In the case of V705~Cas, \citet{2005MNRAS.360.1483E} reported $\tau_{\rm max} \sim 5.5$, attributed to an optically thick dust shell, while for V339~Del a value of $\tau_{\rm max} \sim 0.7$ was found \citep{evans2017rise}.

A possible quiescence counterpart of the nova is a faint source detected in the \emph{Gaia} post-outburst observations in DR2 and DR3.
The \emph{Gaia} measurements correspond to the mean epochs of 2015.5 for DR2 \citep{2018A&A...616A...1G} and 2016.0 for DR3 \citep{2021A&A...649A...1G}. The most recent SMARTS observation corresponds to epoch $\sim$2018.3, with magnitudes $I = 17.52 \pm 0.01$, $R = 17.83 \pm 0.01$, $V = 18.49 \pm 0.01$, and $B = 18.83 \pm 0.01$. These values indicate that the nova remained sufficiently bright to be detected by the \emph{Gaia} survey, consistent with its presence in the DR2 and DR3 catalogs (see Fig.~\ref{lc_optical}).
Subsequently, the discovery coordinates were refined using precise astrometry using \emph{Gaia} DR3 to RA = $13^{\mathrm h}16^{\mathrm m}36^{\mathrm s}.405$ and Dec = $-67^\circ36'47.931''$, and the source was identified as \emph{Gaia} DR3 5845496055203074560. Our photometric distance is in close agreement with the distance derived from the \emph{Gaia} parallax measurements \citep{2021AJ....161..147B,schaefer2022comprehensive}. The Gaia source has a magnitude of $G = 17.9249$ mag and a color index of 
$(BP - RP)$ = 0.9005 mag. After dereddening using the estimated $E(B-V)$ in the respective Gaia filters, the color becomes $(BP - RP)_0$ = 0.08, and the absolute magnitude is $M_G = 3.10$ mag. The color and magnitude lie within the expected range for novae in the HR diagram for cataclysmic variables presented by \cite{10.1093/mnrasl/slz181} (see their Fig.~2). 

\begin{figure}
    \centering
    \includegraphics[width=1.0\linewidth]{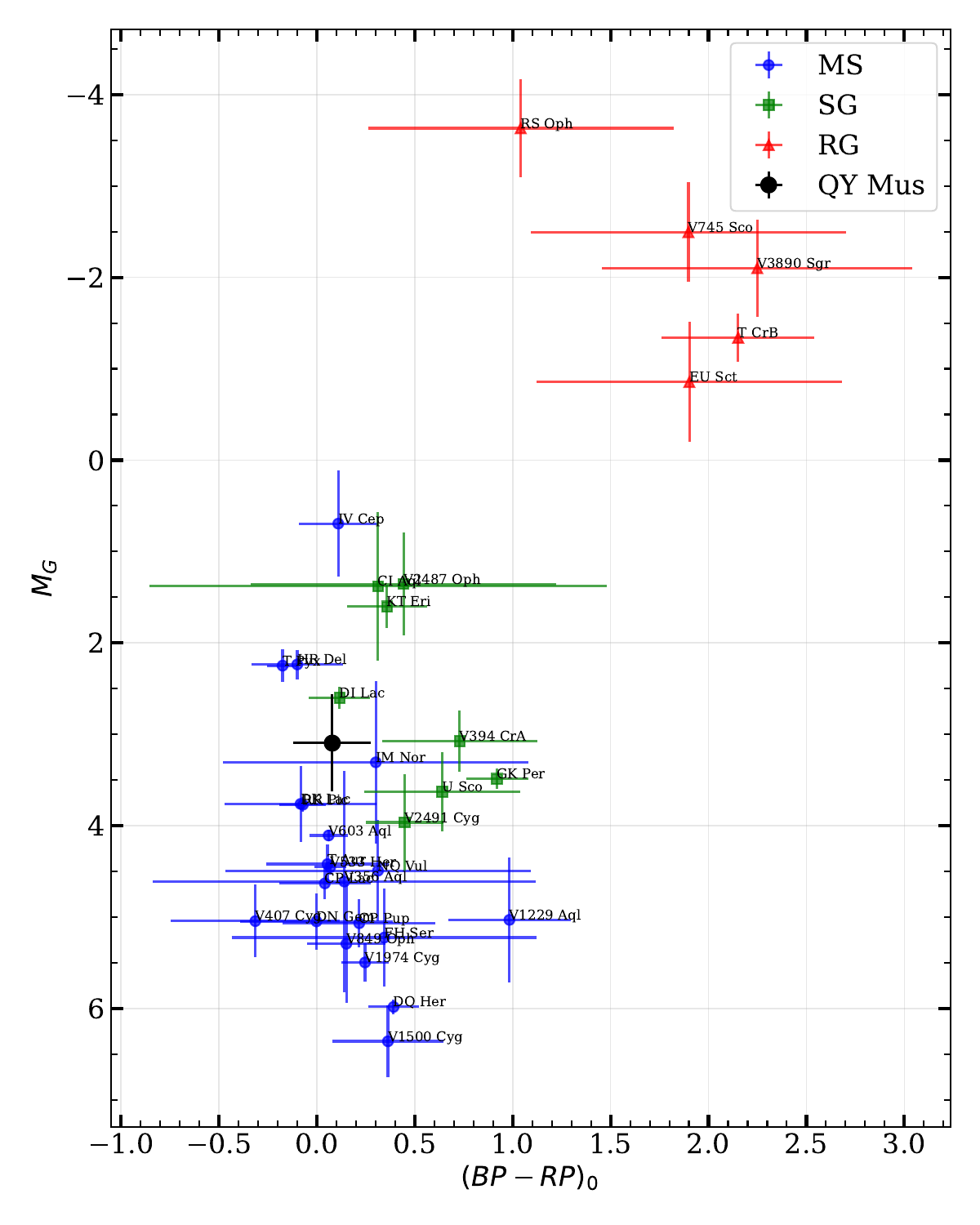}
    \caption{Color--magnitude diagram constructed using Gaia $G$, $BP$, and $RP$ photometry for 34 quiescent novae listed in \citet{2012ApJ...746...61D}, after applying extinction corrections and adopting distances based on Gaia parallaxes \citep{schaefer2022comprehensive}. The position of QY~Mus, derived using Gaia photometry, is overplotted for comparison. }
    \label{gaia_cmd}
\end{figure}

We further compared the Gaia-derived color and magnitude of QY~Mus, together with the distance estimated by \cite{schaefer2022comprehensive} based on Gaia parallaxes, with the quiescent nova population presented by \cite{2012ApJ...746...61D}. While 
\cite{2012ApJ...746...61D} used MMRD-based distances for some novae and ground-based photometry ($B$, $V$, $J$, $H$, $K_s$), our analysis relies on Gaia photometry ($G$, $BP$, $RP$) and parallax-based distance estimates. After applying extinction corrections in the Gaia bands, we constructed a color–magnitude diagram using Gaia data and distances for 34 novae from the quiescent nova sample of \cite{2012ApJ...746...61D}. The extinction values are adopted from \citet{schaefer2022comprehensive}. The uncertainties in color $(BP - RP)_0$ and magnitude $M_G$ are calculated using error propagation as described in \citet{2025A&A...696A.243G}.
We find that QY~Mus occupies a region similar to other novae (see Fig. \ref{gaia_cmd}), consistent with systems hosting main-sequence and subgiant secondary stars; however, the position of QY~Mus in the CMD will be more precisely constrained with upcoming Gaia DR4 and DR5 data releases. The position of a quiescent nova system in such a diagram provides information about the nature of its secondary star \citep{2012ApJ...746...61D}.
With the improved accuracy of Gaia photometry and distances, such color--magnitude diagrams provide a more reliable method for the classification of secondary stars in nova systems. 

Using \emph{TESS} data obtained in 2019 and 2021, \citet{schaefer2022comprehensive} reported an orbital period of 0.901135~d for the system. This period is consistent with those of novae hosting subgiant secondary, which typically have orbital periods in the range of approximately 0.6--10~d.
In contrast, classical novae with main-sequence secondaries generally have much shorter orbital periods, $P \approx 1.4$--10 hours \citep{1995cvs..book.....W, 1997A&A...322..807D}. The orbital period of $\sim$0.9~d (21.6 hours) is therefore significantly longer and strongly implies an evolved subgiant donor.
Future spectroscopic observations aimed at detecting absorption features from the secondary star, as discussed by \citet{2008ASPC..401...31A}, will be important to further constrain its nature.

A possible progenitor of the nova is a faint source detected in the USNO pre-outburst data.
A positional match is found with the USNO-B1.0 source 0223-0511401, located at RA = $13^{\mathrm h}16^{\mathrm m}36^{\mathrm s}.47280$ and Dec = $-67^\circ36'47.9016''$, at an angular separation of approximately $0.4''$ from the \emph{Gaia} position. The USNO magnitudes of this source were $B_2 = 19.91$ mag and $R_2 = 17.50$ mag, with a mean epoch of observation of 1986.2 \citep{2003AJ....125..984M}. This indicates an outburst amplitude of $\sim$10~mag.

\begin{figure}
    \centering
    \includegraphics[width=0.95\linewidth]{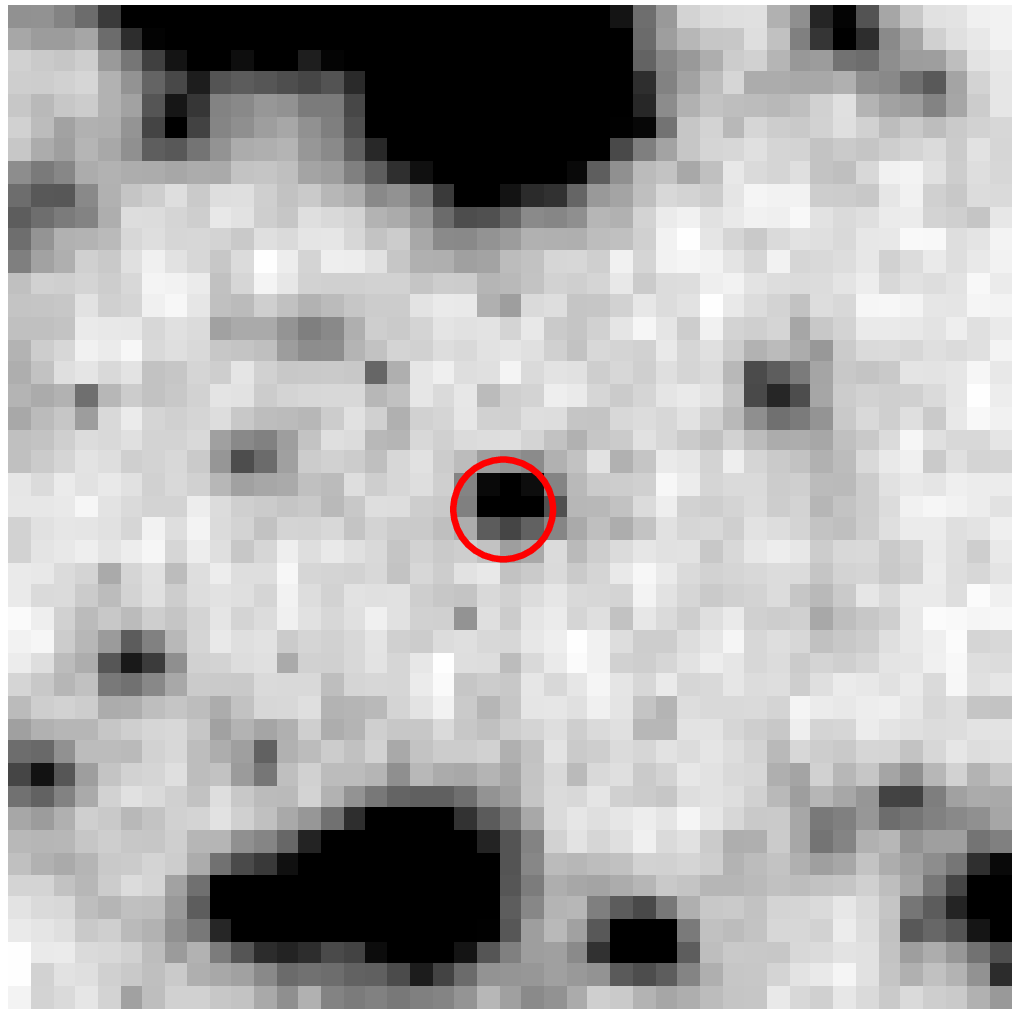}
    \caption{Image from the SuperCOSMOS
H$\alpha$ Survey cropped around the position of
QY Mus. The size is 30x30 arcsec. 
The red circle has a radius
of 1.5 arcsec, and it is centred at the accurate position of Gaia DR3 5845496055203074560 (G=17.92 mag).}
    \label{shs_image}
\end{figure}

Additional pre-outburst imaging is available from the SuperCOSMOS
H$\alpha$ Survey (SHS; \citealt{2005MNRAS.362..689P}), conducted with the UK
Schmidt Telescope (UKST). A wide-field SHS H$\alpha$+[N\,\textsc{ii}] image
covering the nova position was downloaded from the SHS archive\footnote{\href{http://www-wfau.roe.ac.uk/sss/halpha/hapixel.html}{http://www-wfau.roe.ac.uk/sss/halpha/hapixel.html}}. The image,
obtained on 1999 June 21, has a pixel scale of
0.677~arcsec~pixel$^{-1}$ and shows a faint source spatially coincident
with the \emph{Gaia} and USNO positions. The SHS image, shown in the Fig.~\ref{shs_image}, further supports the identification of the progenitor system.

\section{SUMMARY AND CONCLUSIONS}\label{summary}

QY~Mus is a slow nova that erupted in 2008 September. We presented a comprehensive photometric dataset, compiled from multiple public archives, spanning from day 11 to day 3482, covering the maximum, early decline, dust formation, final decline, and quiescence. The spectroscopic evolution is examined using optical spectra obtained between days 94 and 1348. The main findings of this study are summarized and interpreted below.

\begin{enumerate}

\item QY~Mus is a slow nova with $t_2 = 87$ days, suggesting a relatively low-mass CO white dwarf of $\sim 0.7\,M_{\odot}$. Using recent MMRD relations, we estimate a peak absolute magnitude of $M_V = -6.55$ and an outburst luminosity of $\sim 4.24 \times 10^{4}\,L_{\odot}$.

\item The increased decline rate and subsequent recovery in the optical light curve indicate dust formation in the ejecta, with the onset occurring at $\sim$123 days after discovery. This is further supported by the near-infrared light curves and colors. During the dust-dip phase, the position in the color--magnitude diagram deviates from the earlier evolutionary track. The maximum optical depth of the dust is estimated to be $\tau_{\mathrm{max}} \sim 3.2$ around day 146. Based on the light-curve morphology, QY~Mus is classified as a D(137) class nova.

\item The optical color evolution during the dust-minimum phase reflects variations driven by dust extinction. Following the end of the dust phase (after $\sim$200 days), the colors show a transition consistent with the onset of the nebular phase. In particular, $V-R$, $V-I$, and $R-I$ become progressively bluer, while $B-V$ continues to redden. This behavior is in agreement with the spectroscopic evolution, where strong nebular emission lines begin to dominate over the optical continuum, significantly influencing the observed broadband colors.

\item The reddening toward QY~Mus is estimated using intrinsic nova colors at optical maximum and at $t_2$, yielding a mean value of $E(B - V) = 0.59 \pm 0.05$, consistent with independent dust maps. This corresponds to an interstellar extinction of $A_V = 1.84$ mag. Using the distance modulus, we derive a distance of $d = 4.64 \pm 1.16$ kpc, in good agreement with Gaia-based estimates.

\item The spectroscopic observations cover the early decline and nebular phases of the nova evolution. The early decline spectra show prominent hydrogen Balmer and Fe\,\textsc{ii} lines with P~Cygni profiles, consistent with the classification of QY~Mus as an Fe\,\textsc{ii} class nova.
Around day $\sim$233 , it enters the nebular phase, as indicated by the appearance of prominent [O\,\textsc{iii}] lines. The subsequent appearance of high-ionization lines such as [Fe\,\textsc{v}], [Fe\,\textsc{vi}], [Fe\,\textsc{viii}], and [Ne\,\textsc{v}] indicates the evolution of the nova into a higher ionization state. The line profiles of Balmer, [O\,\textsc{i}], and [N\,\textsc{ii}] suggest a non-spherical geometry, consistent with a bipolar or equatorial ring-like structure of the ejecta

\item 
The physical conditions of the ionizing source and the elemental abundances in the ejecta were derived using the \textsc{Cloudy} photoionization code. In total, 41 lines in the day 590 spectrum were fitted in the modeling. We determine the temperature of the central ionizing source to be $(7.08 \pm 0.20)\times10^{5}$~K.
The abundance analysis shows enhancements in nitrogen, oxygen, and neon relative to solar values. However, the neon abundance is lower than that typically observed in neon novae, indicating that QY~Mus is not a neon nova.

\item Mid-infrared observations from \emph{WISE} show significant excess emission at longer wavelengths, especially in the W3 and W4 bands, indicative of emission from cool dust with a temperature of $\sim$400~K.

\item A likely progenitor of QY~Mus is identified through positional coincidence with sources in pre-outburst USNO-B1.0 data and post-outburst \emph{Gaia} DR3 observations. The Gaia counterpart has $G = 17.92$ mag and, after extinction correction, $(BP - RP)_0$ = 0.08 and $M_G = 3.10$ mag, placing it in a region of the color--magnitude diagram consistent with quiescent novae hosting main-sequence or subgiant secondaries. This is further supported by comparison with 34 quiescent novae, where QY~Mus occupies a similar position. The inferred orbital period of 0.9~d is also consistent with systems hosting subgiant companions. A positional match with a USNO-B1.0 source implies an outburst amplitude of $\sim$10 mag. Additional support comes from pre-outburst SuperCOSMOS H$\alpha$ imaging, which shows a faint source coincident with the nova position, reinforcing the identification of the progenitor system.

\end{enumerate}
This research emphasizes the significance of long-term, multiwavelength observations in the understanding of the complicated interplay between the evolution of ejecta, the development of dust, and the binary system in classical novae.

\section*{Acknowledgements}
The authors thank the referee for the thorough evaluation and valuable feedback, which significantly improved the manuscript.
We thank Prof.\ B.\ E.\ Schaefer for valuable discussions on \emph{TESS} data.
This work utilizes data from the AAVSO Database and the SMARTS Atlas, for which we acknowledge their valuable contributions.
We also acknowledge VSOLJ and ASAS databases for providing photometric data used in this work. 

\section*{Data Availability} 
All photometric data used in this work are archival and publicly available. Photoionization modeling was performed using the \texttt{CLOUDY} code, which is freely available from the official website at \url{https://www.nublado.org/}.
The observed optical spectra are available at \url{http://www.astro.sunysb.edu/fwalter/SMARTS/NovaAtlas/}. 

\bibliographystyle{mnras}
\setlength{\bibsep}{0.0pt}
\footnotesize{\bibliography{QY_Mus}}

\label{lastpage}
\end{document}